%% file: mnras_template.tex
\documentclass[fleqn,usenatbib]{mnras}
\usepackage{natbib}
\usepackage{amsmath}

\newcommand{\logg}{\ensuremath{\log g}}

\newcommand{\feh}{\ensuremath{\protect\rm [Fe/H]}}

\newcommand{\teff}{T$_{\rm eff}$}

\usepackage{newtxtext,newtxmath}

\usepackage[T1]{fontenc}
\usepackage{ae,aecompl}
\usepackage{gensymb}
\usepackage{lastpage}

\usepackage{graphicx}	
\usepackage{amsmath}	

\title[]{The 3D kinematics of stellar substructures in the periphery of the Large Magellanic Cloud}
\author[Navarrete et al.]{Camila Navarrete$^{1,2}$\thanks{E-mail: Camila.Navarrete@eso.org}\thanks{Visiting astronomer, Cerro Tololo Inter-American Observatory, NSF’s NOIRLab, which is managed by the Association of Universities for Research in Astronomy (AURA) under a cooperative agreement with the National Science Foundation.}, David S. Aguado$^{3,4,12}$, Vasily Belokurov$^{5,6}$, Denis Erkal$^{7}$, \newauthor 
Alis Deason$^{8,9}$, Lara Cullinane$^{10}$ \& Julio Carballo-Bello$^{11}\footnotemark[2]$
\\
\\
$^{1}$ESO - European Southern Observatory, Alonso de Cordova 3107, Vitacura, Santiago, Chile\\
$^{2}$Université C\^{o}te d'Azur, Observatoire de la C\^{o}te d’Azur, CNRS, Laboratoire Lagrange, Bd de l'Observatoire, CS 34229, 06304 Nice cedex 4, France\\
$^{3}$Instituto de Astrof\'{\i}sica de Canarias,
              V\'{\i}a L\'actea, 38205 La Laguna, Tenerife, Spain\\
$^{4}$Universidad de La Laguna, Departamento de Astrof\'{\i}sica, 
             38206 La Laguna, Tenerife, Spain \\
$^{5}$Institute of Astronomy, University of Cambridge, Madingley Road, Cambridge CB3 0HA, UK \\
$^{6}$Center for Computational Astrophysics, Flatiron Institute, 162 5th Avenue, New York, NY 10010, USA \\
$^{7}$Department of Physics, University of Surrey, Guildford GU2 7XH, UK \\
$^{8}$Institute for Computational Cosmology, Department of Physics, Durham University, South Road, Durham DH1 3LE, U.K \\
$^{9}$Centre for Extragalactic Astronomy, Department of Physics, Durham University, South Road, Durham DH1 3LE, UK \\
$^{10}$Department of Physics and Astronomy, Johns Hopkins University, Baltimore, MD 21218, USA \\
$^{11}$Instituto de Alta Investigaci\'on, Sede Esmeralda, Universidad de Tarapac\'a, Av. Luis Emilio Recabarren 2477, Iquique, Chile\\
$^{12}$Dipartimento di Fisica e Astrofisica, Univerisit\`{a} degli Studi di Firenze, via G. Sansone 1, I-50019 Sesto Fiorentino, Italy\\
}
\date{Accepted XXX. Received YYY; in original form ZZZ}

\pubyear{2023}

\begin{document}

\label{firstpage}
\pagerange{\pageref{firstpage}--\pageref{lastpage}}
\maketitle

\begin{abstract}
We report the 3D kinematics of 27 Mira-like stars in the northern, eastern and southern periphery of the Large Magellanic Cloud (LMC), based on {\it Gaia} proper motions and a dedicated spectroscopic follow-up. Low-resolution spectra were obtained for more than 40 Mira-like candidates, selected to trace known substructures in the LMC periphery. Radial velocities and stellar parameters were derived for all stars. {\it Gaia} data release 3 astrometry and photometry were used to discard outliers, derive periods for those stars with available light curves, and determine their photometric chemical types. The 3D motion of the stars in the reference frame of the LMC revealed that most of the stars, in all directions, have velocities consistent with being part of the LMC disk population, out of equilibrium in the radial and vertical directions. A suite of N-body simulations was used to constrain the most likely past interaction history between the Clouds given the phase-space distribution of our targets. Model realizations in which the Small Magellanic Cloud (SMC) had three pericentric passages around the LMC best resemble the observations. The interaction history of those model realizations has a recent SMC pericentric passage ($\sim$320 Myr ago), preceded by an SMC crossing of the LMC disk at $\sim$0.97 Gyr ago, having a radial crossing distance of only $\sim$4.5 kpc. The previous disk crossing of the SMC was found to occur at $\sim$1.78 Gyr ago, with a much larger radial crossing distance of $\sim$10 kpc.
\end{abstract}




\section{Introduction}
The Magellanic Family is the poster child for binary dwarf galaxy collisions. Thanks to the advanced stage and overt intensity of the Clouds' interaction, we have a rare opportunity to directly observe a variety of phenomena normally postulated to accompany and induce galaxy transformations. This includes tidal stripping and torquing, ram pressure gas removal, dynamical friction, and merger-induced star formation. 

Several pieces of evidence in their morphologies reflect the intense past interaction history of this system. This includes, for example, the Large Magellanic Cloud (LMC) off-set stellar bar \citep{Zhao00, Choi18}, truncation of its outer stellar disk \citep{mackey18}, and the presence of warps \citep{Olsen02, Choi18}. The past orbit of the Clouds suggests that they have experienced a recent close pericentric passage $\leq$250 Myr ago, consistent with the expected formation time of the Magellanic Bridge \citep{Choi18}. Besides this `direct' collision (impact parameter of $\lesssim$ 10 kpc), the details of previous interactions are not yet fully constrained. 

In the periphery of the LMC, numerous stellar substructures -- in the form of arms, clumps and streams -- have been found thanks to wide-field deep-photometric surveys, as well as astrometric data from the \textit{Gaia} mission. This plethora of substructure  reflects its complex interaction history with the Small Magellanic Cloud (SMC) and/or the Milky Way. 
Towards the north, a thin and long stellar stream was first presented in \cite{mac16}, being subsequently studied in detail by several studies 
\citep[e.g.,][]{BelokurovErkal19, Gaia2021, Dalal21, Cullinane22}. Towards the east, a diffuse and extended stellar overdensity has been recovered based on different stellar tracers, receiving different names \citep[e.g., `Eastern Substructure 1', `Eastern Substructure 2'][]{Dalal21}. 
Two thin stellar streams have been recovered towards the southern vicinity of the LMC, most likely embedded in a larger, diffuse stellar substructure \citep{BelokurovErkal19, Dalal21}. 
The origin of these different stellar substructures around the Clouds is not yet established, although most of them are though to be part of the disturbed outer LMC disk based on the properties 
of their stellar populations and in-plane velocities. 

Radial velocities, and with them, full 6D phase-space information for the stars in these substructures, is of paramount importance to understand their motions and to assess their possible origin. Up to date, a limited number of works have collected spectroscopic information for this purpose. In \cite{Cullinane2020}, the \textit{Magellanic Edge Survey (MagES)} is described, which collects multi-object spectroscopy of red giant branch and red clump stars. The \textit{MagES} survey has analysed the northern arm \citep{Cullinane2020}, outer LMC stellar population 
\citep{Cullinane22b} and the SMC outskirts \citep{Cullinane2023}. Similarly, in \cite{Cheng22} the results from APOGEE-2 observations in six fields towards the north and south of the LMC were presented.

While these previous observations were based on red giant branch and/or red clump stars, the errors on the proper motions of these stars tend to be larger than for the most luminous tracers. 
Moreover, at fields at larger distances from the LMC centre, contamination from Milky Way foreground stars can be non negligible. More luminous and less abundant tracers are thus well-suited to uncover the diffuse substructures in the LMC outskirts, as their contamination rate is smaller. Particularly, \cite{deason17} recovered several of the well-known stellar substructures around the LMC based on Mira-candidate stars. Being $\approx$ 3.5 mag  brighter than red clump stars, moderate exposures times are require to derive reliable radial velocities. Motivated by this, we carried out a spectroscopic follow of more than 40 Mira-like stars in the vicinity of the Clouds, aiming to recover their phase-space information and use them to constrain the past interaction history of the Clouds.

This paper is organised as follows: in Section \ref{sec:selection}, 
a summary of the target selection for the spectroscopic follow-up 
is described. In Section \ref{sec:observations}, observations and 
data reduction are discussed. In Section \ref{sec:analisys}, 
we present the spectroscopic analysis of our sample. Additional parameters such as parallax, periods, and heliocentric distances are discussed in Section~\ref{sec:phase-space}. 3D cylindrical velocities in the LMC reference frame for the stars in the immediate LMC periphery are derived in Section~\ref{sec:discussion}. Finally, Section~\ref{sec:sims} presents a comparison between the observations and N-body simulations, while Section~\ref{sec:conclusions} presents the summary and conclusions of this work.

\section{Target Selection}\label{sec:selection}

Candidate Mira variable stars were selected from \textit{Gaia} DR1 data following the procedure outlined in \cite{deason17}. In that work, repeated observations of sources during the initial phase of the \textit{Gaia} mission 
were used to identify stars that show signs of variability. In particular, the \textit{Gaia} `variability amplitude' was used to identify stars that 
show signs of variability: $A = \sqrt{N_{\rm obs}}\sigma(F)/F$, where, 
$N_{\rm obs}$ is the number of CCD crossings, and $F$ and $\sigma(F)$ are 
the flux and flux error, respectively (see also \citealt{belokurov17}). 
This variability information was combined with infrared photometry from 
the Two Micron All Sky Survey \citep[2MASS;][]{Skrutskie06} and the 
Wide field Infrared Survey Explorer \citep[WISE;][]{Wright10} 
to select candidate giant stars in the vicinity of the LMC. 
For more details, please see Section~2 of \cite{deason17}. 

\begin{figure}
    \centering
    \includegraphics[width=0.49\textwidth]{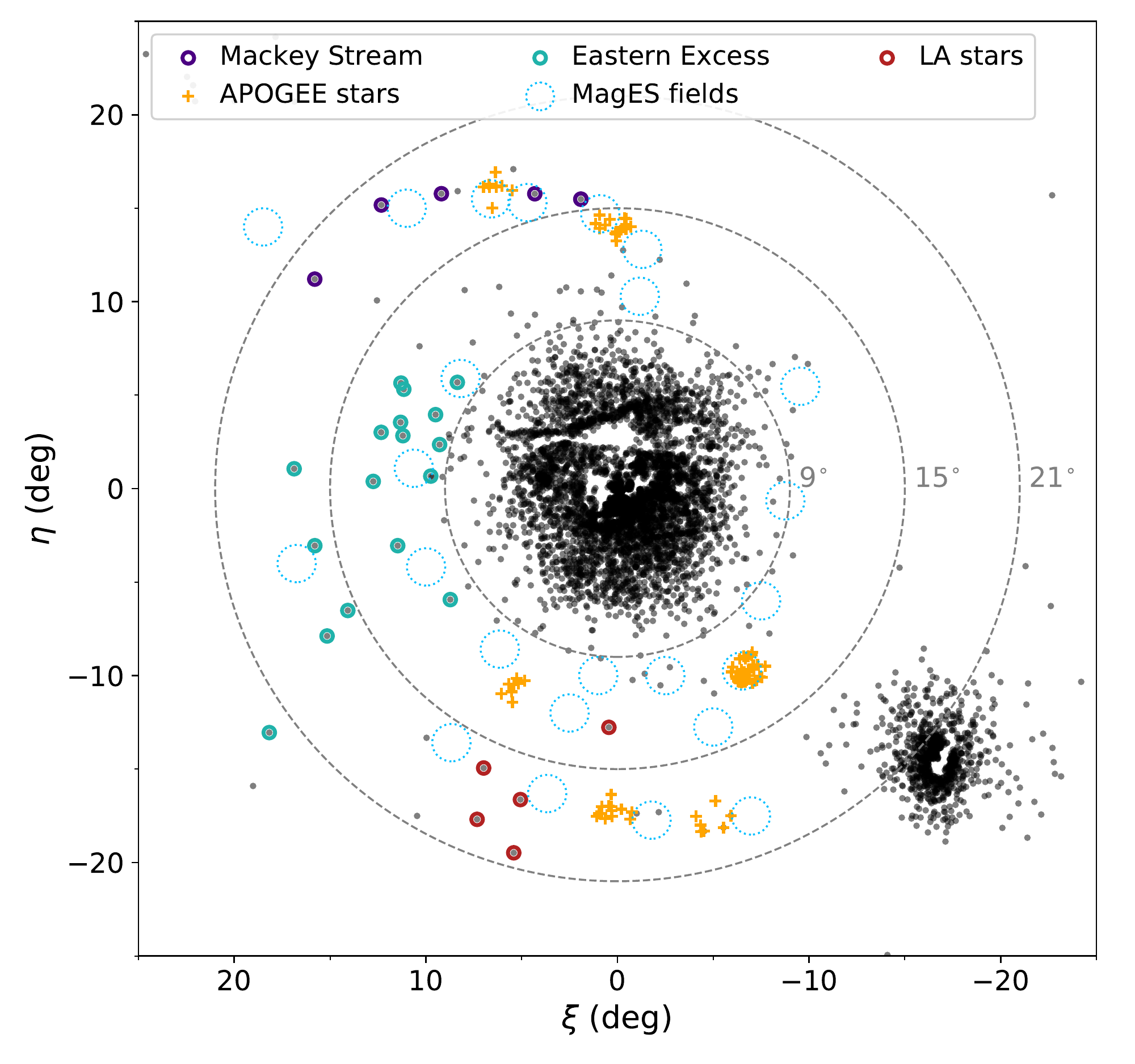}
    \caption{Mira candidates in the vicinity of the LMC and SMC,  
    selected following the cuts detailed in \protect\cite{deason17}. 
    The coordinates ($\eta$, $\xi$) are in the gnomonic (tangent-plane) 
    projection. The dashed circles correspond to 9, 15 and 21{\degree} of angular 
    separation from the LMC center. The targets selected 
    for follow-up observations are marked with coloured circles, including 
    five stars in the so-called Mackey Stream, 17 stars from the Eastern 
    excess group, and five stars in the LA group. Blue unfilled circles mark 
    the position of the fields in \protect\cite{Cullinane22b} while yellow crosses 
    are individual stars from \protect\cite{Cheng22}.}
    \label{fig:Mira_targets}
\end{figure}

\begin{figure}
    \centering
    \includegraphics[width=0.49\textwidth]{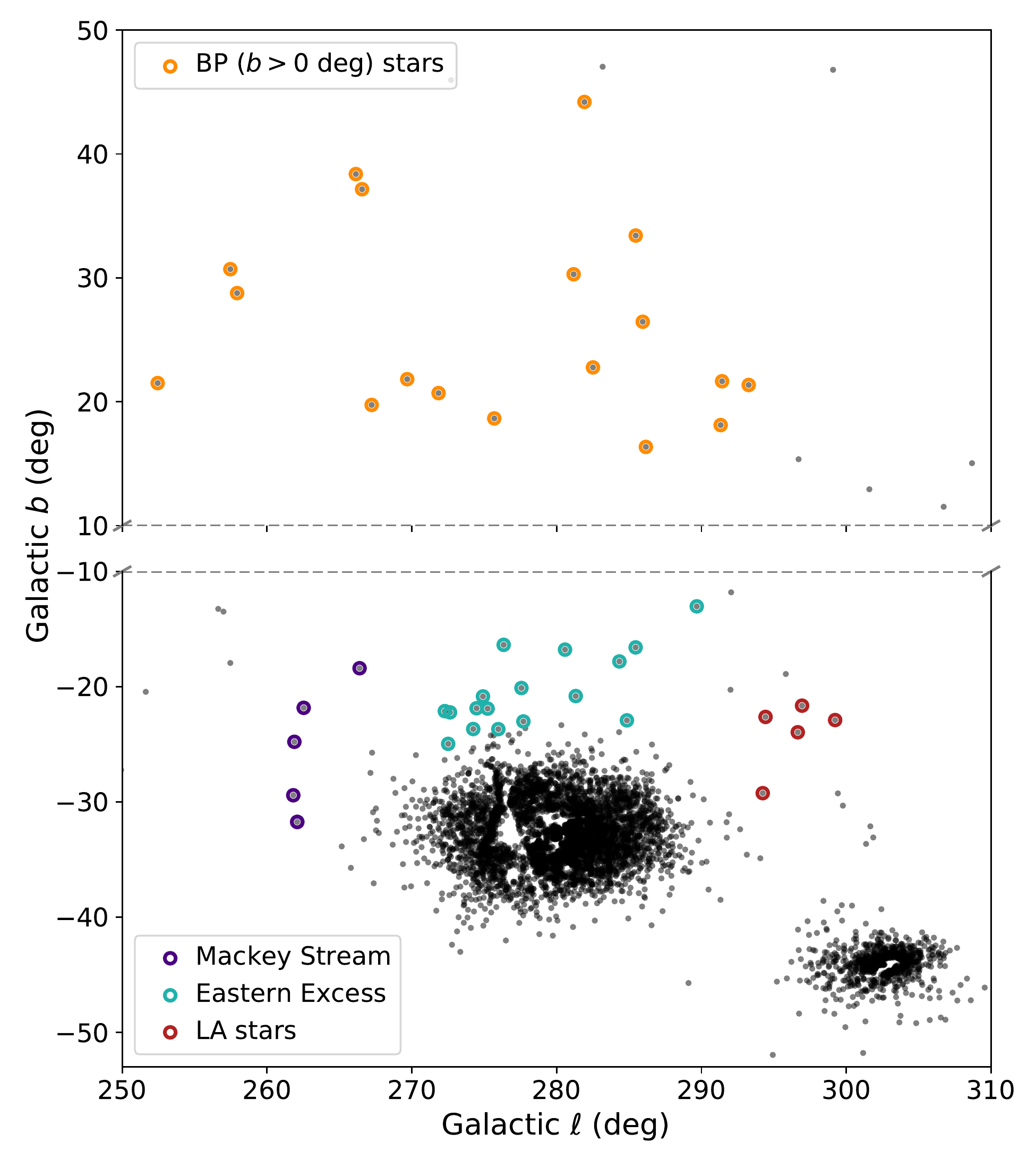}
    \caption{Galactic coordinate distribution of our target Mira stars. 
    The three groups in the vicinity of the LMC and SMC (`Mackey Stream', 
    `Eastern excess' and `LA') are the same targets as shown in 
    Figure~\ref{fig:Mira_targets}. The 17 target stars above the Galactic 
    plane (`BP' group) are shown as orange open circles.}
    \label{fig:Mira_targets_galactic}
\end{figure}

We select a sub-sample of these giant candidates for low-resolution spectroscopic 
follow-up to obtain radial velocities. In particular, we focus on the periphery of the Clouds, where there are interesting groups of stars to the East (`Eastern excess', EE), North (`Mackey Stream', MS), and South (possibly associated with the Leading Arm, dubbed `LA') of the LMC. 
Figure~\ref{fig:Mira_targets} shows the location of the individual 
Mira-like stars observed, in a gnomonic (tangent-plane) projection, 
where the tangent point is located at the center of the LMC 
($\alpha_0$, $\delta_0$) = (82\fdg25, --69\fdg5). All the Mira candidates from 
\cite{deason17} are shown as grey points. The locations of two previous 
spectroscopic campaigns around the periphery of the Clouds are included for reference (fields from \citealt{Cullinane22} and stars from \citealt{Cheng22}). 
We also targeted some candidate giants away from the Magellanic Clouds (BP), which are above the disk plane (see Fig.~\ref{fig:Mira_targets_galactic}). 
These could potentially be associated with Magellanic debris or distant halo stars with a different origin (see Section~\ref{sec:BPstars} for further discussion).

\section{Observations and Data Reduction}
\label{sec:observations}

Mira candidates in the Magellanic periphery were observed for 
4 nights with the COSMOS spectrograph \citep{Martini14} on the 
4m V\'ictor Blanco telescope at Cerro Tololo Inter-American 
Observatory (CTIO)\footnote{Chilean time proposal CN2017B-0910.}. 
Long-slit spectra for 45 Miras were obtained using 
the Blue VPH Grism covering a wavelength range of 5600-9600 
{\AA} and a 1.0 arcsec slit, reaching a spectral resolution of $\sim$2000. 
Single exposures of 300s to 1800s per target were acquired, as well 
as one or two wavelength calibration arcs with Hg+Ne lamps observed 
at the same airmass as the targets. Radial velocity standard Miras 
\citep[R Lep, HD 75021;][]{Menzies06} were also observed each night, 
with exposure times of 0.5 to 5 seconds.

A standard data reduction (flat-fielding, bias subtraction, extraction 
and wavelength calibration) was performed with the \emph{onespec} package 
in IRAF\footnote{IRAF is distributed by the National Optical Astronomy 
Observatory, which is operated by the Association of Universities for 
Research in Astronomy (AURA) under cooperative agreement with the National 
Science Foundation.} \citep{tod93}. The lamps used were the non-standard 
Hg+Ne more suitable for calibration in the red part of the spectrum.
A barycentric correction is then performed for each individual exposure, 
considering the Julian date, the place of the observatory on the 
surface of the Earth and the coordinates of each object.

\begin{figure*}
\centering
	\includegraphics[width=0.92\textwidth, angle=180]{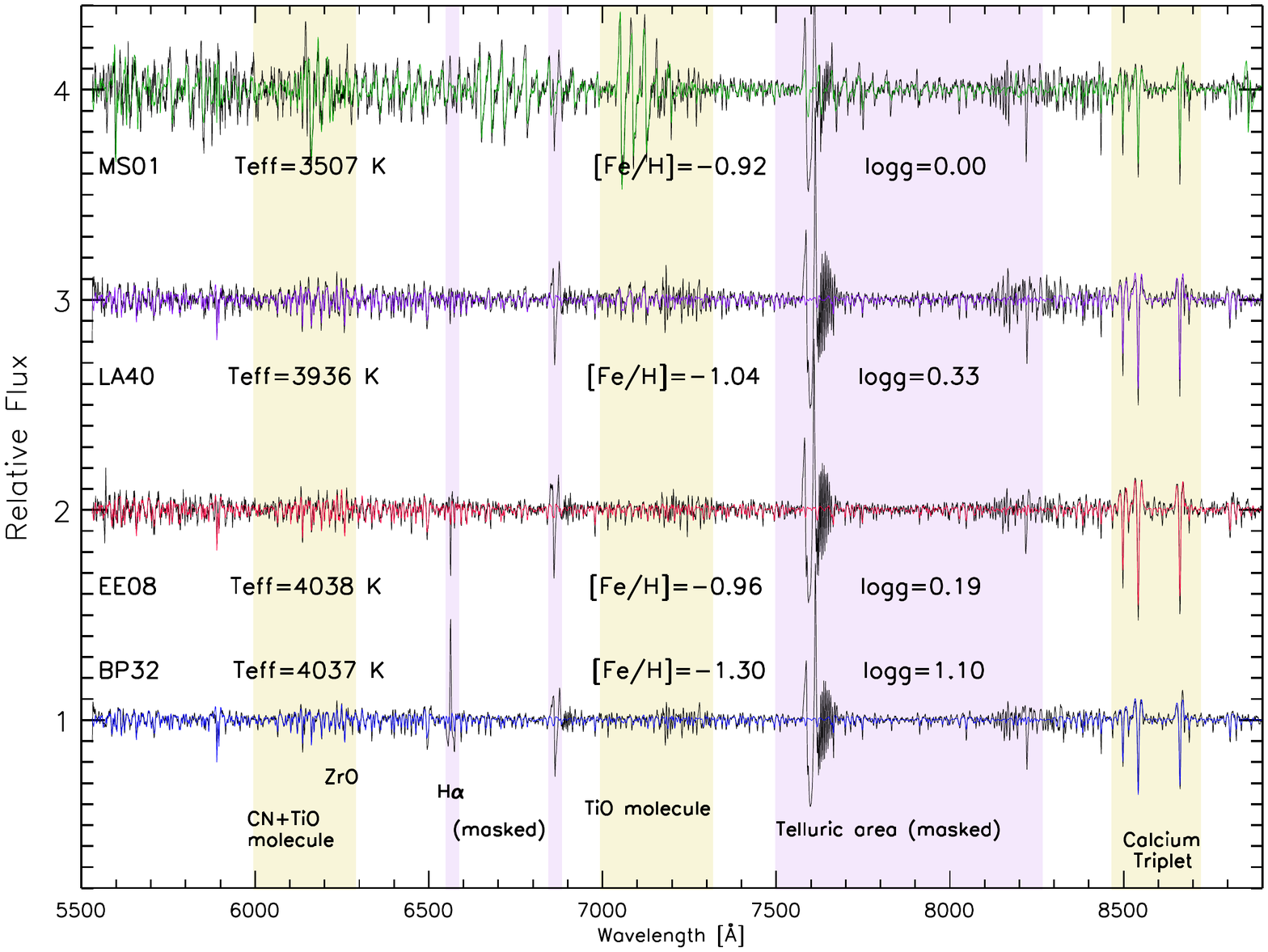}
        \caption{COSMOS spectra for a subsample of the observed targets together with the best fit derived with FERRE. The main stellar parameters derived for each target are also displayed. Yellow shadowed regions are for relevant regions (e.g., location of the Calcium triplet), while purple shadowed areas are those masked and not used for the FERRE fit.}
    \label{temp_spec}
\end{figure*}

\input{table1_2dec.tex}
\section{Spectroscopic Analysis}\label{sec:analisys}

The spectra of Mira variables change significantly during their pulsation cycle; therefore stellar parameters (brightness, opacity and surface 
temperature) are phase-dependant. In addition, shock waves in the 
pulsating atmospheres of Mira variables can produce emission lines. 
In particular, close to the maximum brightness, they show very strong 
H$_\alpha$ emission lines \citep{joy1926}. Several objects in our sample 
(13 out of 45) show these emission lines (see Fig.~\ref{temp_spec}). 

\subsection{Radial Velocity estimates}\label{sec:analisys_rv}

To derive radial velocities (V$_{\rm rad}$), we first corrected 
the spectra for barycentric motion calculated with the {\tt rvcorrect} 
package. We then use {\tt fxcor} we to calculate the cross-correlation function (CCF) based on the algorithm 
by \citet{tonry79}, using a template computed with the ASS$\epsilon$T 
code \citep{koe08} by assuming the following stellar parameters: 
T$_{\rm eff}=4000$\,K, $\logg =$ 1, and $\rm [Fe/H]=$ --1.0.

Under our observational setup, the COSMOS data cover a region severely 
affected by telluric absorption. We therefore employ the region around 
the \ion{Ca}{ii} triplet (8498, 8542, and 8662 \AA) for cross-correlation assuming the stellar Calcium origin.  
Depending on the quality of the Ca triplet, we 
derived a range of V$_{\rm rad}$ uncertainties up to 80\,km\,s$^{-1}$ 
but the average value is $\sim$15\,km\,s$^{-1}$. The derived radial 
velocities and uncertainties are summarised in Table \ref{tab:stellar_params}. 

We also measured the V$_{\rm rad}$ difference with emission lines (H$_\alpha$) 
when available; the results were in agreement with \citet{Menzies06} 
with reported shifts up to 30\,km\,s$^{-1}$ along the pulsating phase. 
Our derived values are shown in Table \ref{tab:stellar_params}. 
Furthermore, we do not observe Ca in emission in contrast with \citet{gillet85b} 
that reported both 8498 and 8542 {\AA} lines in emission for several stars, 
suggesting our observations were not made close to the maximum of the pulsation cycle. 

\begin{figure}
    \centering
    \includegraphics[width=0.48\textwidth]{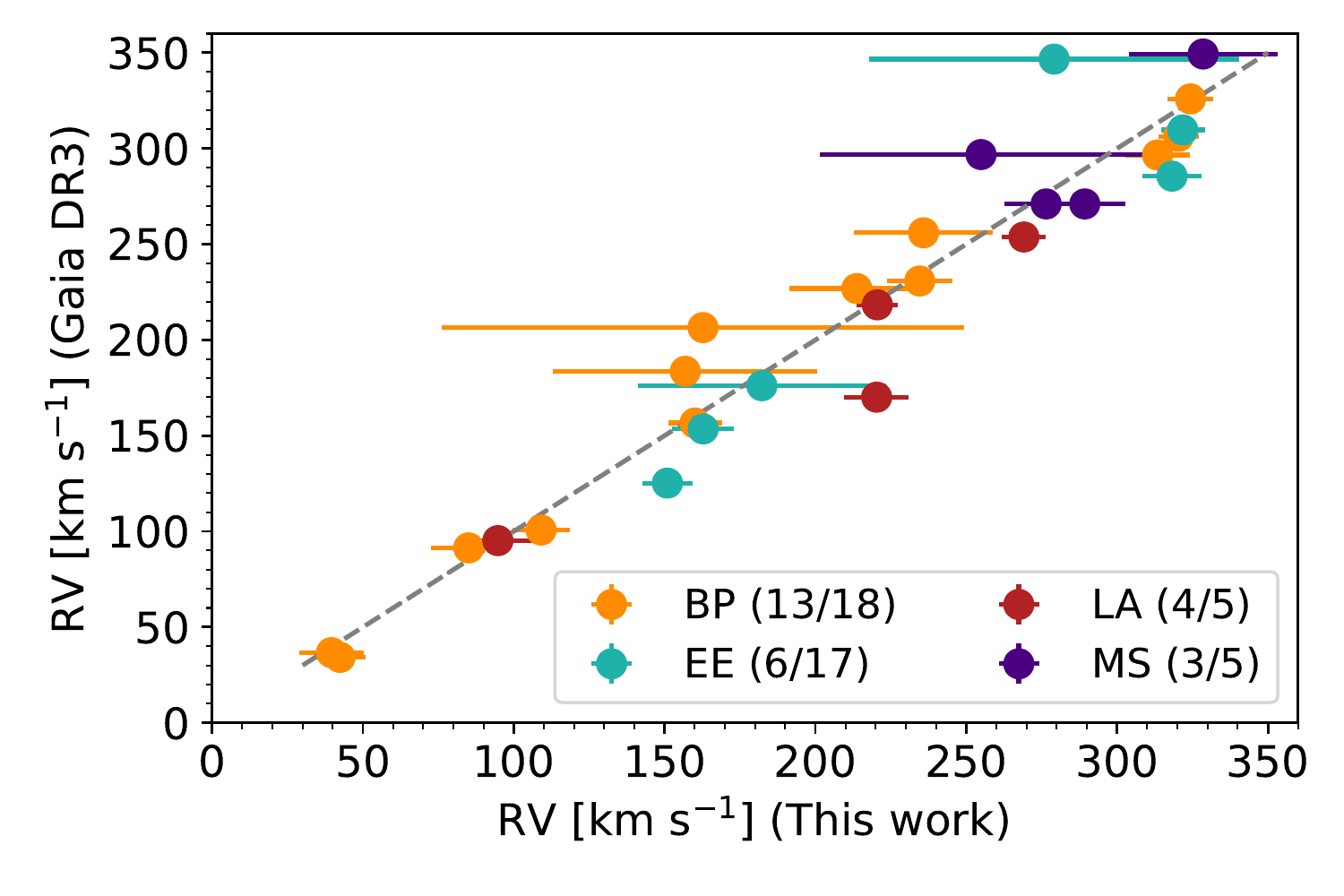}
    \caption{Radial velocity comparison between the values obtained 
    based on COSMOS spectra and {\it Gaia} DR3 main source catalogue. Stars 
    corresponding to the BP, EE, LA and MS groups are shown as orange, 
    green, red and purple circles, respectively. The number in parenthesis 
    in the legend is the number of stars having {\it Gaia} radial velocities relative to
    the total number of stars observed in this work.}
    \label{fig:rv_comp}
\end{figure}

At the time of our observations, only {\it Gaia} DR1 was available, without 
any measurement of radial velocities. Nowadays, the {\it Gaia} DR3 \citep{Gaia22} main source 
catalogue contains radial velocity measurements for more than 33 
million sources with magnitudes G$_{\rm RVS} <$ 14 mag. 
Table~\ref{tab:Gaia_RV} presents the radial velocities 
reported in the {\it Gaia} DR3 main source catalogue, after cross-matching it 
with our target stars, using a 1.0 arcsec matching radius. 26 out of the 45 
stars observed with COSMOS have radial velocities reported in {\it Gaia}. 
There is an overall good agreement between those values and the ones 
derived based on the COSMOS spectra (see Fig.~\ref{fig:rv_comp}). This 
agreement reinforces our assumption that the possible radial velocity 
variations due to the pulsation cycle are of the order of magnitude of 
the measured uncertainties. As almost 20 of our targets do not have radial 
velocities reported in {\it Gaia}, and those that have measurements available are 
in good agreement, the following analysis will use the COSMOS radial velocities.

\subsection{Stellar Parameters}
 
Assuming they may slightly vary along the phase, we derive stellar 
parameters by spectroscopic fitting with the FERRE code\footnote{{\tt FERRE} 
is available from http://github.com/callendeprieto/ferre} \citep{alle06}. 
FERRE is able to employ different algorithms to minimise the $\chi^{2}$ 
against a library of stellar spectra \citep{alle18}. Then, by interpolating 
within the nodes of the grid, the code provides the best stellar model and 
the most likely effective temperature ($T_{\rm eff}$), surface gravity 
($\log g$) and overall metallicity ([Fe/H]).

Both the data and the models were normalised with a running mean filter 
of 30 pixels. For the purpose of spectroscopic analysis, we masked the 
Balmer-$\alpha$ line and the regions most affected by telluric lines. Thus, the 
code does not take into consideration the $\chi^{2}$ from those areas. In 
Fig.~\ref{temp_spec} we show a sub-sample of nine spectra and the best fit 
derived with FERRE. Relevant areas are shaded in yellow while the masked 
regions are coloured in purple.

In Table~\ref{tab:stellar_params}, we also provide a flag quality by visual inspection of each 
FERRE fit. In some cases, due to the relatively low quality of the spectra and/or the presence of significant telluric lines, even outside of our masked area, it was not possible 
to obtain a good fit. For these cases, we set {\tt flag=0};  for the remainder we set {\tt flag=1}. Additionally, we performed a complementary 
analysis deriving metallicities for the entire sample only considering the 
information contained in the region of the Calcium triple around 8400-8700\,\AA. 
We fixed {\teff} and {\logg} from the FERRE analysis, and then  calculated 
\feh$_{\rm CaT}$ masking everything outside of the considered region. The results 
are summarised in Table \ref{tab:stellar_params}, and in most cases are in 
good agreement with the {\feh}. However, in some cases, we find a significant deviation between the two determinations, up to 1 dex. The majority of these cases correspond to 
objects with {\tt flag=0}, indicating the fit is poor. However, there are 7 
cases -- BP34, EE10, EE17, EE21, LA42, LA44, and MS04 -- where the fit appears 
reasonable but the metallicities are incompatible. Those objects are quite cool 
(\teff$<3800$\,K) and the combination of strong molecular bands and the presence 
of telluric lines could explain the difference. For those objects, we recommend 
considering the overall metallicity \feh as only a tentative value. 

Finally, BP30 and MS04 have been observed twice; in both cases we derived compatible 
stellar parameters, \teff, \logg, and \feh. The \feh$_{\rm CaT}$ for these two 
objects is clearly incorrect and could also be explained as a result of their low-temperature. 
In fact, the \feh$_{\rm CaT}$ for BP30 is located at the limit of the grid, 
indicating the code did not find metallicity information within the CaT.

\begin{table}
	\centering
	\caption{Radial velocities from {\it Gaia} DR3. Radial velocity, its error, and S/N of the spectra are the columns \texttt{radial\_velocity}, \texttt{radial\_velocity\_error} and \texttt{rv\_expected\_sig\_to\_noise}, respectively.}
	\label{tab:Gaia_RV}
\begin{tabular}{rccc}
\hline
ID     & RV$_{\rm Gaia}$ & RV error      & S/N  \\
       & (km s$^{-1}$)   & (km s$^{-1}$) &           \\
\hline
BP22   & 306.33	& 2.15	& 11.6 \\
BP24   & 325.88 & 3.13	&  7.6 \\ 
BP25   &  36.64	& 3.22	&  8.1 \\
BP26   & 100.68	& 1.81	& 10.7 \\
BP27   & 296.57	& 1.89	& 11.9 \\
BP28   & 183.59	& 1.75	&  6.1 \\
BP32   & 230.77	& 3.14	&  7.4 \\
BP33   &  34.17	& 2.99	&  7.9 \\
BP35   & 226.80	& 1.34	& 11.4 \\
BP36   & 206.43	& 3.65	& 10.8 \\
BP37   & 255.94	& 2.21	& 10.9 \\
BP38   &  91.36	& 2.43	&  8.4 \\
BP39   & 156.59	& 4.12	&  8.0 \\
EE06   & 309.69	& 4.95	&  6.4 \\
EE10   & 285.53	& 3.69	&  4.5 \\
EE16   & 346.68	& 2.15	&  5.5 \\
EE17   & 176.06	& 3.9	&  8.0 \\
EE18   & 153.48	& 3.47	&  4.5 \\
EE52   & 125.20	& 4.87	&  3.8 \\
LA41   & 218.32 & 7.91	&  3.5 \\
LA42   & 253.76	& 4.0	&  3.1 \\
LA43   &  95.14	& 7.84	&  5.7 \\
LA44   & 170.08	& 6.88	&  3.7 \\
MS02   & 296.82	& 1.46	&  5.5 \\
MS03   & 349.34	& 2.41	&  3.9 \\
MS04   & 271.02	& 5.15	&  4.9 \\
\hline
\end{tabular}
\end{table}

The {\it Gaia} DR3 parallax, effective temperature, surface gravity, metallicity and 
spectral type for our targets, when available, are presented in Table~\ref{table_Gaia}. 

\section{6D phase-space information}\label{sec:phase-space}

\subsection{Parallaxes and proper motions from Gaia DR3}

Even though most of the observed targets have large radial velocities 
(V$_{\rm rad} \geq$ 200 km s$^{-1}$), it is clear that some targets are 
most likely foreground contaminants (V$_{\rm rad} \leq$ 50 km s$^{-1}$). 

In order to discard those stars and identify potential contaminants 
with large velocities, parallaxes from {\it Gaia} DR3 have been analysed. 
All the targets have parallaxes between --0.051 mas to 0.036 mas, 
with the exception of four stars (BP29, BP30, BP31 and BP34) which have parallaxes 
$>$ 2 mas (see Table~\ref{table_Gaia}). These four stars also have proper motions much larger than the other targets, having absolute values greater 
than 3 mas yr$^{-1}$ in each component. In addition, these stars have the lowest radial velocities of all the targets -- 
from --11.70 $\pm$ 6.12 km s$^{-1}$ (weighted mean of the two 
measurements for BP30) to 27.85 $\pm$ 9.65 km s$^{-1}$ (BP34) -- and the 
highest values of surface gravity (logg $\approx$ 5.0). We therefore exclude these 
four dwarf foreground stars for the remaining analysis.

\begin{figure}
    \centering
    \includegraphics[width=0.48\textwidth]{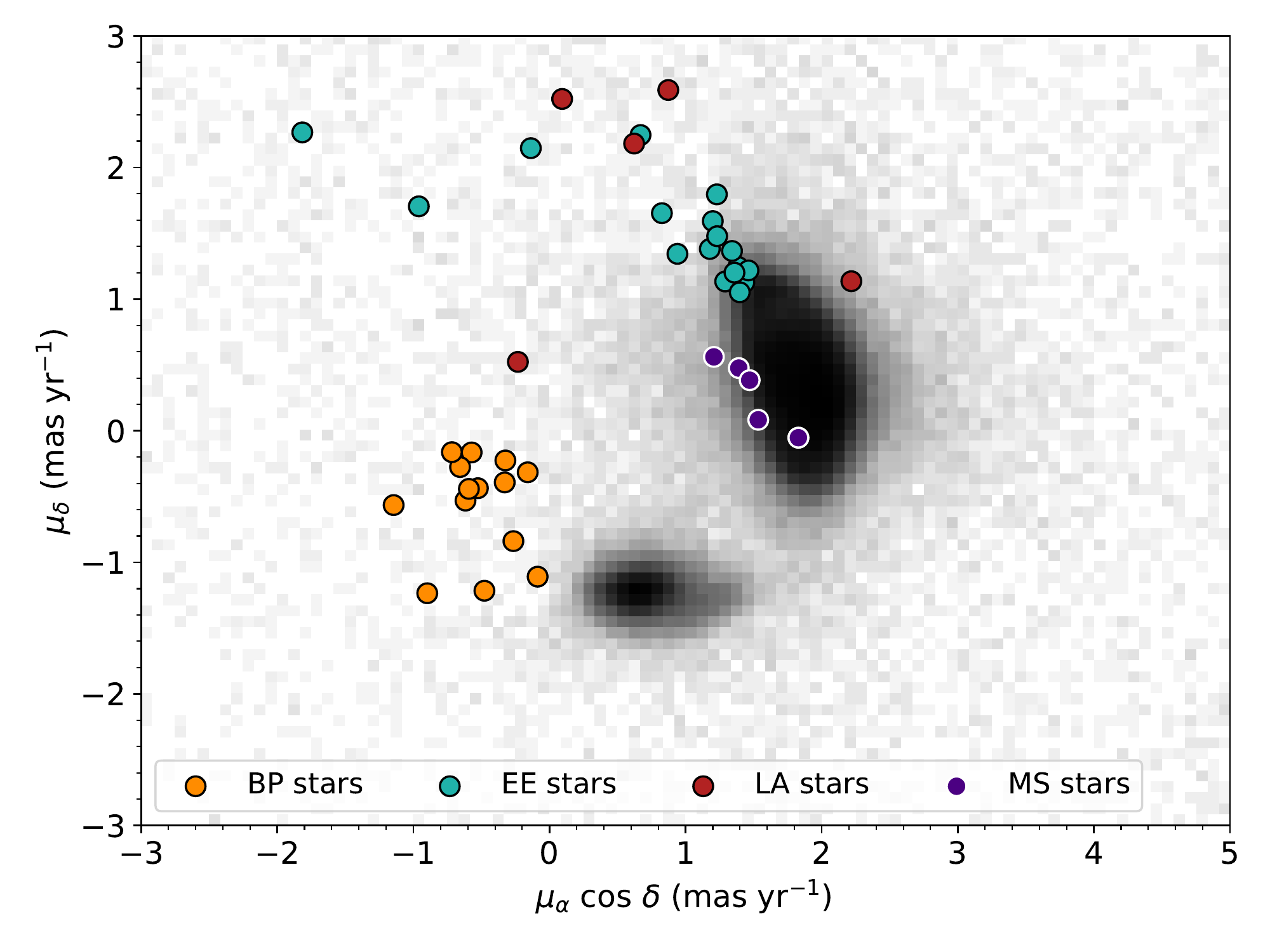}
    \caption{{\it Gaia} DR3 proper motions for stars up to 10 deg from the LMC and 
    SMC centers, brighter than G $<$ 17 mag and redder than (BP-RP) $>$ 1.3 mag. 
    Our targets are shown as coloured circles.}
    \label{fig:ppm}
\end{figure}

Figure~\ref{fig:ppm} shows the proper motion of our sample, excluding the four contaminants. The grey scale corresponds to a sample of 
stars up to 10 deg from the LMC and/or SMC center, having G $<$ 17 mag 
and (G$_{\rm BP}$--G$_{\rm RP}$) $>$ 1.3 mag (most likely RGB stars). 
The LMC and SMC stars are clearly seen as overdensities at 
($\mu_{\alpha}$, $\mu_{\delta}$) = (1.6, 0.5) mas yr$^{-1}$ and (1.1, --1.15) 
mas yr$^{-1}$, respectively. Stars from the MS group have the most similar proper motions to the bulk of the LMC stars, in agreement with previous studies that suggest this northern 
arc is most likely associated with the LMC disk population \citep[e.g.,][]{Cullinane22}. 

Interestingly, the stars belonging to the BP group have proper 
motions somewhat different from those of the SMC population. 
These stars were originally selected as potential debris stars 
stripped from the SMC, based on their positions in the sky and 
predictions from simulations of the SMC's disruption under the LMC potential \citep{deason17}. It is also possible that a large fraction 
of the stars in this group are distant Miras in the Milky Way halo, 
given their brightness and proper motions close to 0 mas yr$^{-1}$. This group of stars is discussed further in Section \ref{sec:BPstars}

After removing the likely foreground contaminants based on their 
parallaxes, proper motions, and radial velocities, our final sample 
consists of 41 Mira candidates, with measurements of radial 
velocities and stellar parameters from COSMOS 
spectra, and parallaxes and proper motions from {\it Gaia} DR3.

In order to study the kinematic behaviour of these stars and 
their possible origin, a distance estimate is needed. However, at the 
distance of the LMC, {\it Gaia} parallaxes are not a reliable 
measurement of distance. In the next section, we discuss 
different approaches to derive the distance of the stars belonging 
to the MS, LA and EE groups (the closest to the LMC in the sky), 
making use of well-known Period-Luminosity relations as well as 
Magellanic RR Lyrae (RRL) stars as anchors for the distance. 

\subsection{Light curves and periods}\label{sec:dist}

\begin{figure*}
 \includegraphics[width=\textwidth]{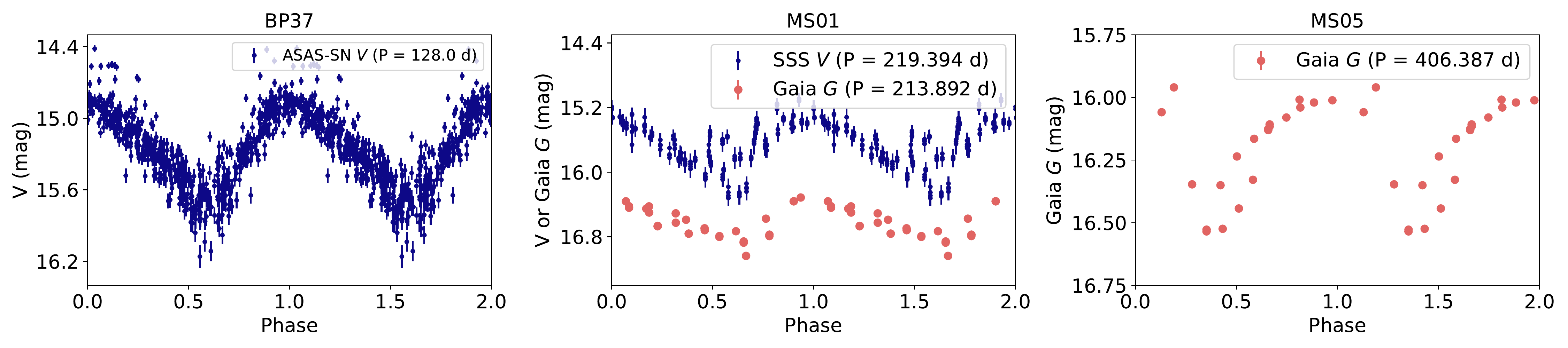}
 \caption{{\it Gaia} $G$ and optical $V$-band light curves for three 
 Mira stars in our sample. From left to right: {\it Gaia} $G$-band light 
 curve and period for MS01; ASAS-SN optical $V$-band light curve 
 and period for BP37; periods and light curves from the {\it Gaia} and 
 CRTS surveys for MS05.}\label{fig:lightcurves_ex}
\end{figure*}

In order to confirm the variable nature of our targets and use their 
periods, if available, to derive distances, we searched for associated light curves 
in various datasets. From the {\it Gaia} DR2 time series, 12 light curves were 
retrieved. However, only 10 of them have reported frequencies in the {\it Gaia} 
{\tt gaiadr2.vari\_long\_period\_variable} table. The corresponding periods 
are reported as P$_{\rm Gaia}$ in Table~\ref{tab:periods}. Light curves from 
the All-Sky Automated Survey for Supernovae \citep[ASAS-SN,][]{Jayasinghe18} 
and from the South Catalina survey \citep[SSS,][]{Drake17} were also recovered 
for 23 variables (6 with light curves and periods also reported in {\it Gaia}). The 
two most significant periods from the CRTS survey were used to derive phase-folded 
light curves, adopting the period that allows us to recover one full pulsation 
cycle. Figure~\ref{fig:lightcurves_ex} shows the {\it Gaia} $G$ or optical $V$-band CRTS 
or ASAS-SN light curves for MS01, BP37 and MS05. The time series have very different 
numbers of epochs; nonetheless, the variability is well recovered. 

We adopted the periods from CRTS when available, instead of the {\it Gaia} 
periods, since the former have a larger number of epochs. When only {\it Gaia} or 
ASAS-SN light curves were available, the corresponding period was adopted. 
The recovered periods range from 65 to 450 days, except for EE13 which has 
a period of only 1.13 days according to the SSS catalogue. However, the light curve 
suffers from considerable scatter and the period reported could be due to a 
1-day alias. Therefore, for the remainder of the paper, we consider that a reliable 
period is not available for this star.

In total, periods were recovered for 27 out of 41 Mira candidates. 
For both the BP and MS group, a high fraction of the Mira candidates 
have reported periods (12 out of 
14 BP stars, and all the MS stars). In the case of the EE and LA stars, light curves were found 
for only 8 (out of 17) and 2 (out of 5) stars respectively. As a result, only for a subsample of the Mira candidates can distances be derived using any of the Period-Luminosity (PL) relations for LMC stars \citep[see e.g.,][]{Soszynski07}. We proceed to separate this subsample into C-rich and O-rich Miras, and fundamental, 
first-overtone and long-secondary period pulsators, to apply the corresponding 
PL relation and derive their distances.

\input{table2.tex}

\subsection{Photometric chemical-types}

Among long-period variables (LPVs), 
different masses and types of stars can be found. Depending on the 
evolutionary state (RGB or AGB star), mass regime and surface chemistry 
(C/O-rich), these stars follow different $K_{\rm S}$ and {\it Gaia} $G$ 
PL relations \citep{Lebzelter19}. The K$_{\rm S}$ PL 
diagram for LPVs in the LMC shows different relations for C-rich and O-rich 
red giants, as well as for fundamental and first overtone pulsating Miras, 
semi-regular variables and long-secondary period variables 
\citep[see e.g.,][]{Soszynski07}. Therefore, identifying the sub-type 
of each of the targets in our sample is necessary in order to derive 
distances for those stars with measured periods (see Section~\ref{sec:dist}).

A powerful diagram to separate the different sub-types of pulsating 
giant stars is the (W$_{\rm RP}$ - W$_{\rm J, Ks}$) versus K$_{\rm S}$ 
diagram presented by \cite{Lebzelter18}. Combining optical {\it Gaia} 
$G$, G$_{\rm BP}$, G$_{\rm RP}$ magnitudes and near-IR 2MASS $J$ and 
K$_{\rm S}$ magnitudes, the Wesenheit functions W$_{\rm RP}$ and 
W$_{\rm J, Ks}$ allow us to use reddening-free magnitudes. \cite{Lebzelter18} 
defined a boundary to separate C-rich from O-rich LPVs in the LMC based 
on the (J-K$_{\rm S}$) colours, with C-rich stars being those with redder 
near-IR colour (i.e., having W$_{\rm RP}$ - W$_{\rm J, Ks}$ $\gtrsim$ 0.8 mag). 
Among the O-rich stars, using population synthesis models, 
\cite{Lebzelter18, Lebzelter19} define different boundaries to separate 
low, intermediate and high-mass pulsating AGB stars, while the faintest 
O-rich stars (K$_{\rm S} \geq$ 12 mag) are mostly early AGB and RGB 
stars from the tip of the RGB in the LMC. 

We cross-match our targets with the {\it Gaia} DR3 and 2MASS catalogues, 
using a search radius of 1.0 arcsec. The Wesenheit functions were 
calculated following the original definition in \cite{Lebzelter18}, as follows:

\begin{align}
W_{\rm BP, RP} &= G_{\rm RP} - 1.3 \left( G_{\rm BP} - G_{\rm RP} \right) \\
W_{\rm J, Ks} &= K_{\rm S} - 0.686 \left( J - K_{\rm S} \right)
\end{align}

\begin{figure}
  \centering
	\includegraphics[width=0.5\textwidth]{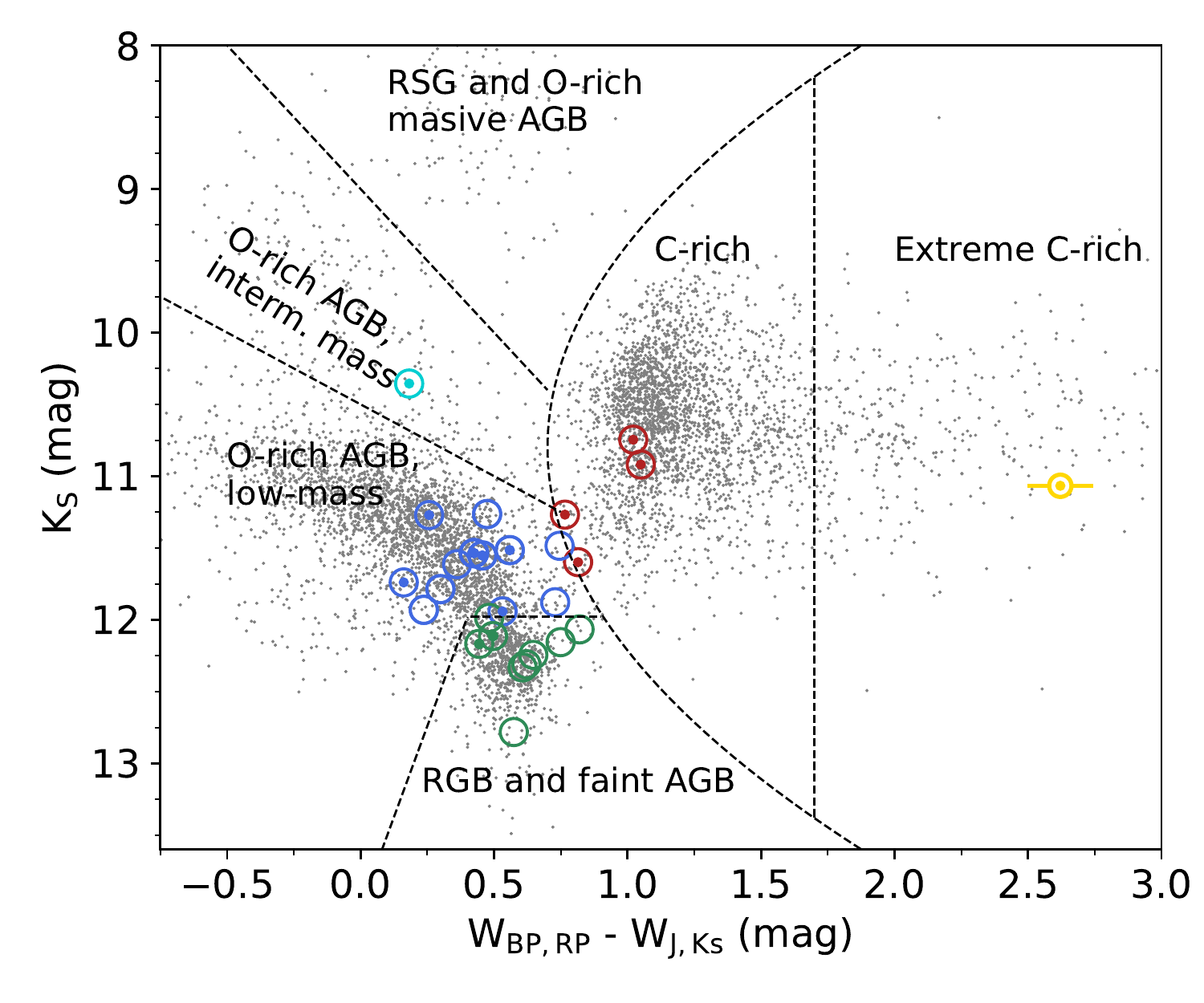}
     \caption{W$_{\rm BP,RP} -$ W$_{\rm J, K_{\rm S}}$ versus 
     K$_{\rm S}$ diagram for LPV stars around 
     the LMC (grey dots) as well as the Miras-like stars belonging to the 
     EE, MS and LA groups. The boundary lines that separate C-rich 
     from O-rich stars, as well as the different mass regimes, are marked as dashed lines (taken from 
     \protect\citealt{Lebzelter19}). 
     Our sample is mainly comprised of O-rich low-mass AGB 
     stars. Those stars with periods available in the literature 
     are denoted with a coloured dot inside the marker.}
    \label{mira_subtypes}
\end{figure}

Figure~\ref{mira_subtypes} shows the location of our target stars 
(from the three groups close to the LMC: MS, EE and LA) in the 
(W$_{\rm RP}$ - W$_{\rm J, Ks}$), K$_{\rm S}$ diagram. The boundaries 
from \cite{Lebzelter19} are shown as dashed lines, while the 
corresponding sub-types are shown as green (O-rich RGB and faint, 
early AGB stars), blue (O-rich low-mass AGB stars), cyan (O-rich 
intermediate-mass AGB stars), red (C-rich) and yellow (extreme 
C-rich) circles. None of our targets have colours and magnitudes in 
the location of red supergiants (RSG) or O-rich massive AGB stars. 
We find that most (22 out of 
27) of our targets are O-rich stars, while five are C-rich  including one extreme C-rich star (though note that two stars have colours on the boundary between C and O-rich 
groups). Among the O-rich stars, 
9 have colours and magnitudes consistent with RGB stars or 
faint, early AGB (green circles); one star (cyan circle) has 
properties consistent with being an intermediate-mass 
(initial stellar masses, M$_{i} \gtrsim$ $\sim$2.0 to 
$\sim$3.2 M$_{\odot}$) AGB star; and the remainder 
(12 out of 27) are consistent with being O-rich low-mass 
(M$_{i} \sim$ 0.9 to $\sim$1.4 M$_{\odot}$) early AGB or 
thermal-pulsating AGB stars. 

Half of the stars (14 out of 27) in Figure~\ref{mira_subtypes} 
have periods available, and distances can thus be estimated using 
PL relations. 

The targets from the BP group were not included in this diagram 
since the boundaries from \cite{Lebzelter18} are defined for the 
LMC distance modulus. The Mira stars in the BP group are far from 
the Clouds and their distances could be quite different from the LMC. 
We discuss this particular group in the next Section.

\subsection{Distance determination}

\subsubsection{Stars around the Large Magellanic Cloud}\label{sec:lmcdist}

\begin{figure}
\centering
 \includegraphics[width=0.48\textwidth]{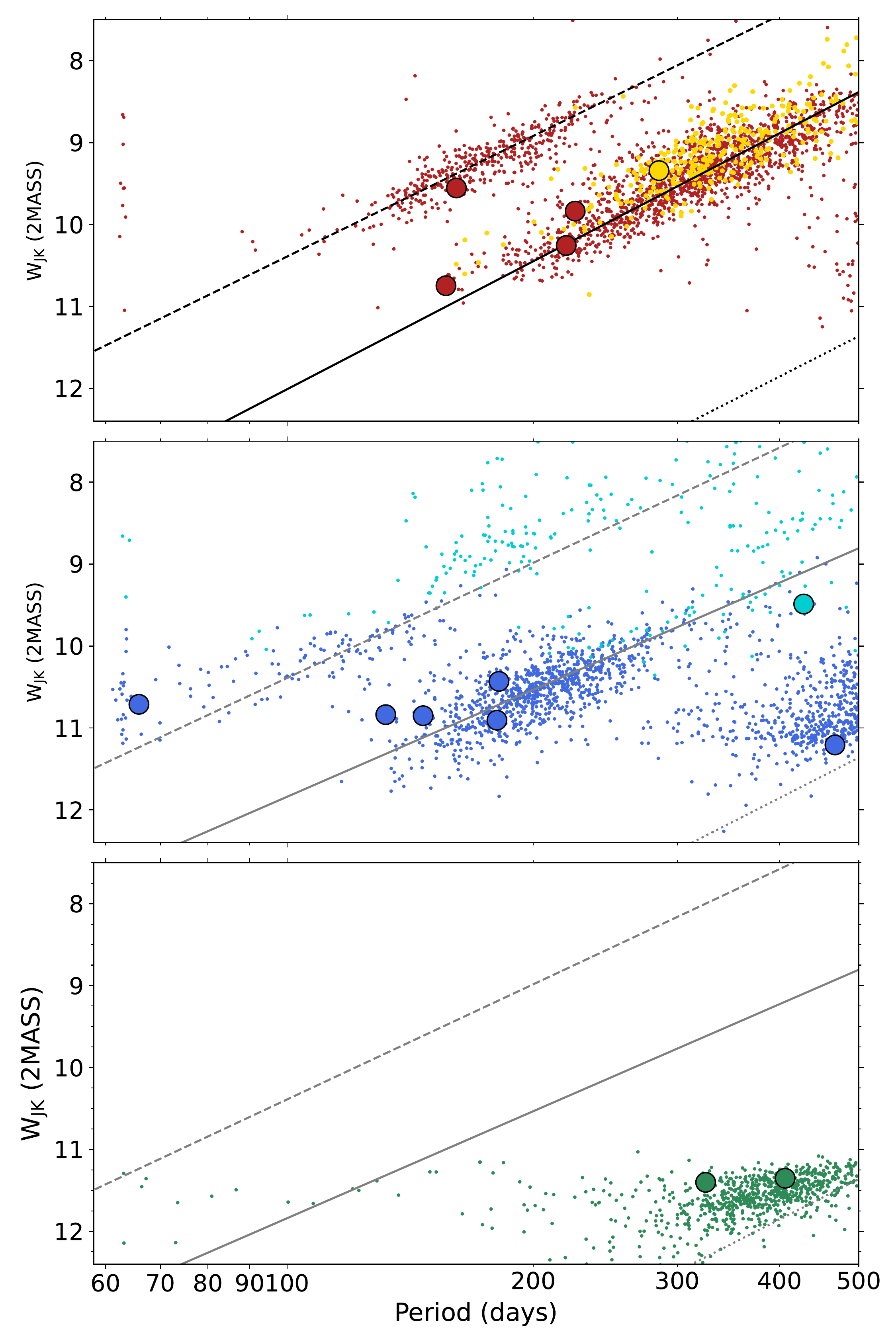}
 \caption{Wesenheit index W$_{\rm J, Ks}$ -- PL diagrams for LPVs in 
 the LMC, colour-coded according to their classification in the 
 {\it Gaia}-2MASS diagram (see Fig.~\ref{mira_subtypes}). In the top 
 panel, C-rich (red) and extreme C-rich (yellow) stars are shown, 
 along with the five Mira variables found to be consistent with 
 this classification. The middle panel shows those stars classified 
 as O-rich AGB stars of low (blue) and intermediate (cyan) mass, 
 while the bottom panel presents the RGB and faint AGB stars (green). 
 The best-fit PL relations for the first overtone, fundamental mode 
 and long-secondary period sequences are shown as dashed, solid and 
 dotted grey/black lines for O/C-rich stars, according to 
 \protect\cite{Soszynski07}.}\label{fig:PL_relations}
\end{figure}

Following the classification into sub-types from \cite{Lebzelter19}, 
we divided the Mira candidate stars around the LMC from {\it Gaia} into C-rich stars, 
O-rich stars and RGB or faint AGB stars (with magnitudes consistent 
with the tip of the LMC's AGB). The top panel of Figure~\ref{fig:PL_relations} 
shows the Wesenheit W$_{\rm J,Ks}$ - period diagram for those stars classified 
as extreme C-rich (yellow points) and C-rich (red points); the middle 
panel shows the O-rich low (blue) and intermediate-mass (cyan) AGB 
stars, and the bottom panel includes only the stars consistent with 
being RGB or faint AGB stars (green points). The larger coloured points 
correspond to our targets (one extreme C-rich, four C-rich, six O-rich 
low-mass, one O-rich intermediate-mass and two RGB/faint AGB stars), 
while the smaller coloured dots are the LPV stars from the {\it Gaia DR2} catalogue 
of LPV candidates in the different subgroups shown in  Fig.~\ref{mira_subtypes}.

In each panel, the best-fit PL relations from \cite{Soszynski07}, 
based on the Optical Gravitational Lensing Experiment (OGLE) observations, are shown for first-overtone pulsators 
(dashed lines), fundamental mode pulsators (solid lines) and long-secondary 
period variables (dotted lines), for both O-rich (grey) and C-rich (black) 
red giant variables. The coloured dots in each panel correspond to LPVs 
variables in the LMC, selected following the same selection criteria as 
in \cite{Lebzelter18}, and cross-matched with the 2MASS photometry. 

As already noted by \cite{Lebzelter19}, there is 
significant scatter and deviations from the best PL relations in all pulsation modes, reflecting inaccurate 2MASS mean magnitudes 
(one single measurement over the pulsation period) as well as 
underestimated periods, shifting the distribution of LMC LPVs towards 
shorter periods. This effect is clearly seen in the case of the 
RGB and faint AGB stars (bottom panel in Figure~\ref{fig:PL_relations}), 
which are found mostly pulsating in the long-secondary period 
(the stars with the longest periods in the sample), and which are 
systematically offset from the PL relation. In fact, all the stars 
with the longest ($>$300 days) periods in our sample -- i.e., EE19, 
MS01 (C-rich RGB and faint AGB stars), LA40 (low-mass AGB star), and 
EE17 (intermediate-mass AGB star) -- are systematically offset from the 
PL relation of the most likely pulsation mode. Therefore, these 
LPVs cannot be used as precise distance estimators 
unless the time-series photometry is sufficient to recover reliable 
period estimates.

In the case of the extreme C-rich stars (yellow points, middle panel 
of Figure~\ref{fig:PL_relations}), these are expected to be pulsating 
in the fundamental mode. Nonetheless, there is considerable scatter in 
their magnitudes with respect to the corresponding PL sequence 
derived by \cite{Soszynski07}. This scatter can be due to stars that 
are highly reddened and/or stars producing high-opacity dust grains.

If all our targets were of the Mira variability class, they should 
be pulsating in the fundamental mode. Nonetheless, given the selection 
made to recover them, we cannot rule out that some of our targets are 
instead semi-regular variables which could be pulsating in the fundamental 
or first-overtone pulsation modes. In fact, as Fig.~\ref{fig:PL_relations} 
shows, C-rich stars with large periods (P $> 300$ d) tend to be in the locus 
for those stars in the long-secondary period group (green dots), while stars 
with short periods ($P <$ 100 d) preferentially pulsate in the first-overtone 
mode. Assuming that each of our targets is pulsating in the pulsation mode 
corresponding to the closest relation in Fig.~\ref{fig:PL_relations}, 
we derived their distance moduli using the Wesenheit W$_{\rm J,Ks}$ PL relations from 
\cite{Soszynski07}. To obtain the individual distances, a distance modulus 
of 18.477 $\pm$ 0.004 (statistical) $\pm$ 0.026 (systematic) mag 
\citep{pietrzynski19} for the LMC was adopted. 

In the case that all our targets are Miras, pulsating in the fundamental mode, 
the distances would be $\sim$2 times shorter ($\sim$ 4 times larger) for those 
stars considered to be pulsating in the first-overtone (long-secondary period). 
This scenario is unlikely, however, as our targets are close to the LMC in the 
sky and it is more plausible to find distant ($\sim$ 50 kpc) long-period (Mira-like) 
variables around the LMC periphery than finding Milky Way halo Miras at less than 
30 kpc close to the LMC, or Mira candidates at distances more than 200 kpc. 

\begin{figure}
\centering
 \includegraphics[width=0.48\textwidth]{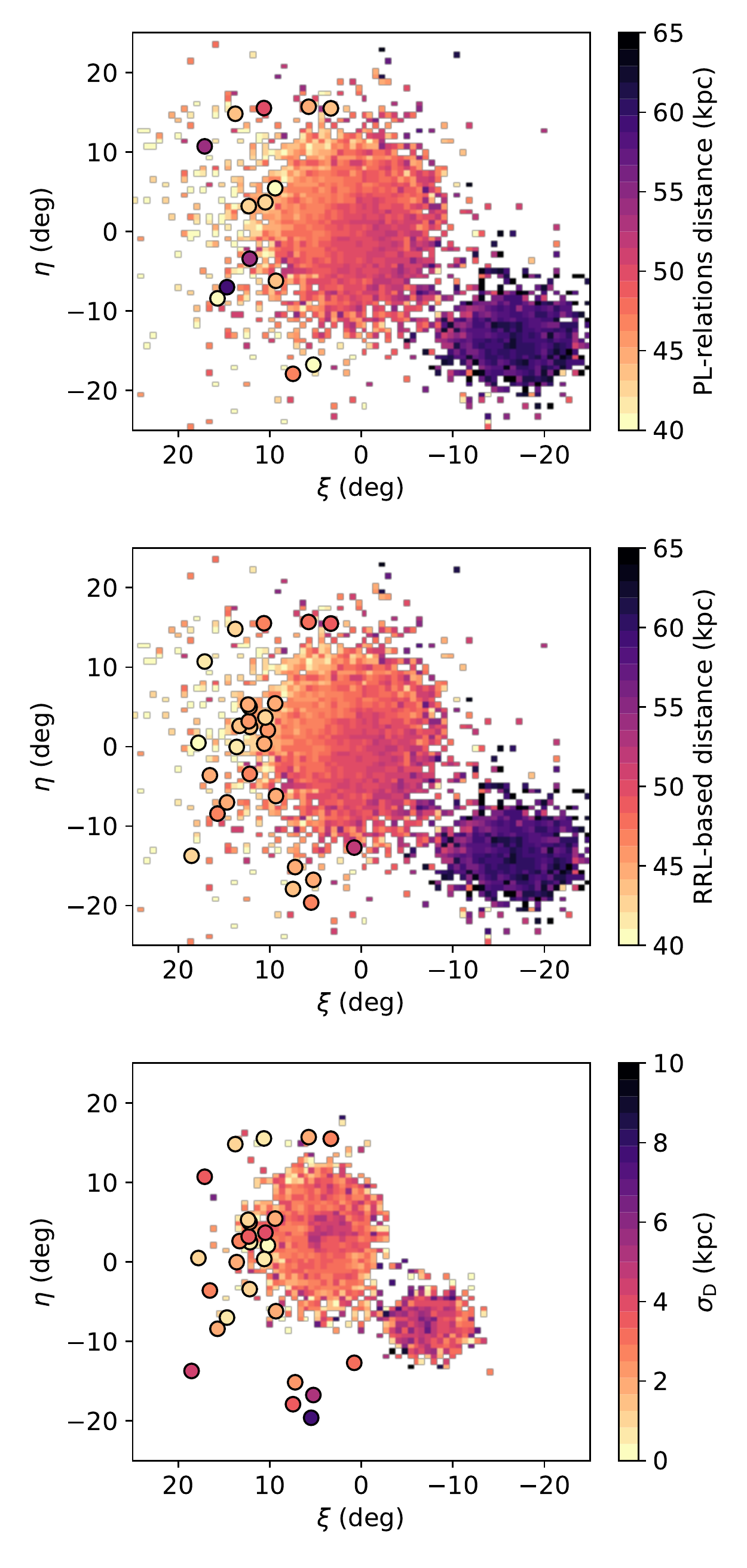}
 \caption{Heliocentric distances for different stellar tracers around the Clouds. 
 The 2D histogram shows the mean distance (top and middle panels) of RRL stars around 
 the Clouds and its standard deviation (bottom panel). The top panel shows as coloured 
 circles those targets with available distances, based on PL relations. The middle and 
 bottom panels respectively show the mean heliocentric distances and standard deviation for RRL stars 
 around the on-sky position of our targets.}\label{fig:RRL_distances}
\end{figure}

In order to confirm the order of magnitude of the distances obtained with the limited 
information we have about our targets' pulsation mode, Fig.~\ref{fig:RRL_distances} shows 
the distances obtained for our targets on top of the distribution of the Magellanic 
RRL stars. The top and 
middle panels show the distance of our targets based on the PL relations and the 
median distance of the closest 5 RRL to each of our targets. The bottom 
panel shows the standard deviation in the RRL distance. The figure shows that RRLs trace 
substructures around our targets from 42 to 52 kpc, with a mean distance of 45 kpc 
(with median errors bars of less than 5 kpc). While our targets are at distances from 
36 kpc to 59 kpc, their mean distance is $\sim$ 46 kpc, as in the case of RRLs. 

Table~\ref{tab:Miras_dist} presents the different sub-types for our targets based on Figure~\ref{mira_subtypes} 
(i.e., C-rich or low-mass, intermediate-mass and RGB O-rich stars), their pulsation mode based on 
Figure~\ref{fig:PL_relations}, and their associated distance. For those stars at the boundary of being 
O-rich/C-rich (MS03 and MS05), both possible distances are reported.

\begin{table}
	\centering
	\caption{Heliocentric distances from PL relations. The sub-types (C- or O-rich, low-mass, intermediate-mass or RGB) for those 
	stars with available periods is presented. The pulsation mode (FM: fundamental mode, FO: first-overtone, LSP: long-secondary period) adopted to derive distances is also included. Errors on the distances are from the propagation of uncertainties on the PL relation coefficients, LMC distance modulus and 2MASS magnitudes.}
	\label{tab:Miras_dist}
\begin{tabular}{rccc}
\hline
ID     & Sub-type & Pulsation mode   & Distance  \\
       &          &                  & (kpc)      \\
\hline
EE06   & O-rich (low)        & FM  & 39.8 $\pm$ 1.2  \\
EE07   & O-rich (low)        & FM  & 53.9 $\pm$ 1.7  \\
EE10   & O-rich (low)        & FM  & 43.5 $\pm$ 1.3  \\
EE14   & Extreme C-rich      & FM  & 43.0 $\pm$ 1.9  \\         
EE16   & C-rich              & FM  & 42.3 $\pm$ 1.7  \\   
EE17   & O-rich (mid)        & FM  & 59.3 $\pm$ 2.3  \\
EE19   & O-rich (RGB)        & LSP & 36.7 $\pm$ 1.5  \\
\hline
LA40   & O-rich (low)        & LSP & 47.0 $\pm$ 2.0  \\
LA41   & O-rich (low)        & FO  & 38.9 $\pm$ 1.0  \\
\hline
MS01   & O-rich (RGB)        & LSP & 44.2 $\pm$ 1.8  \\
MS02   & C-rich              & FO  & 53.8 $\pm$ 1.6  \\
MS03   & C-rich/O-rich (low) & FM  & 44.1 $\pm$ 1.6/44.2 $\pm$ 1.3 \\
MS04   & O-rich (low)        & FM  & 43.8 $\pm$ 1.3  \\
MS05   & C-rich/O-rich (low) & FM  & 50.0 $\pm$ 2.1/47.3 $\pm$ 1.6 \\
\hline
\end{tabular}
\end{table}

\subsubsection{Stars above the Galactic plane (BP group)} \label{sec:BPstars}

Among our targets, after discarding the four foreground stars with large parallaxes, there 
are 14 Mira candidates located above the Galactic plane. Given 
their position on the sky, these stars could be unrelated to the Clouds, and distance determination is thus more uncertain. 

\begin{figure}
\centering
 \includegraphics[width=0.48\textwidth]{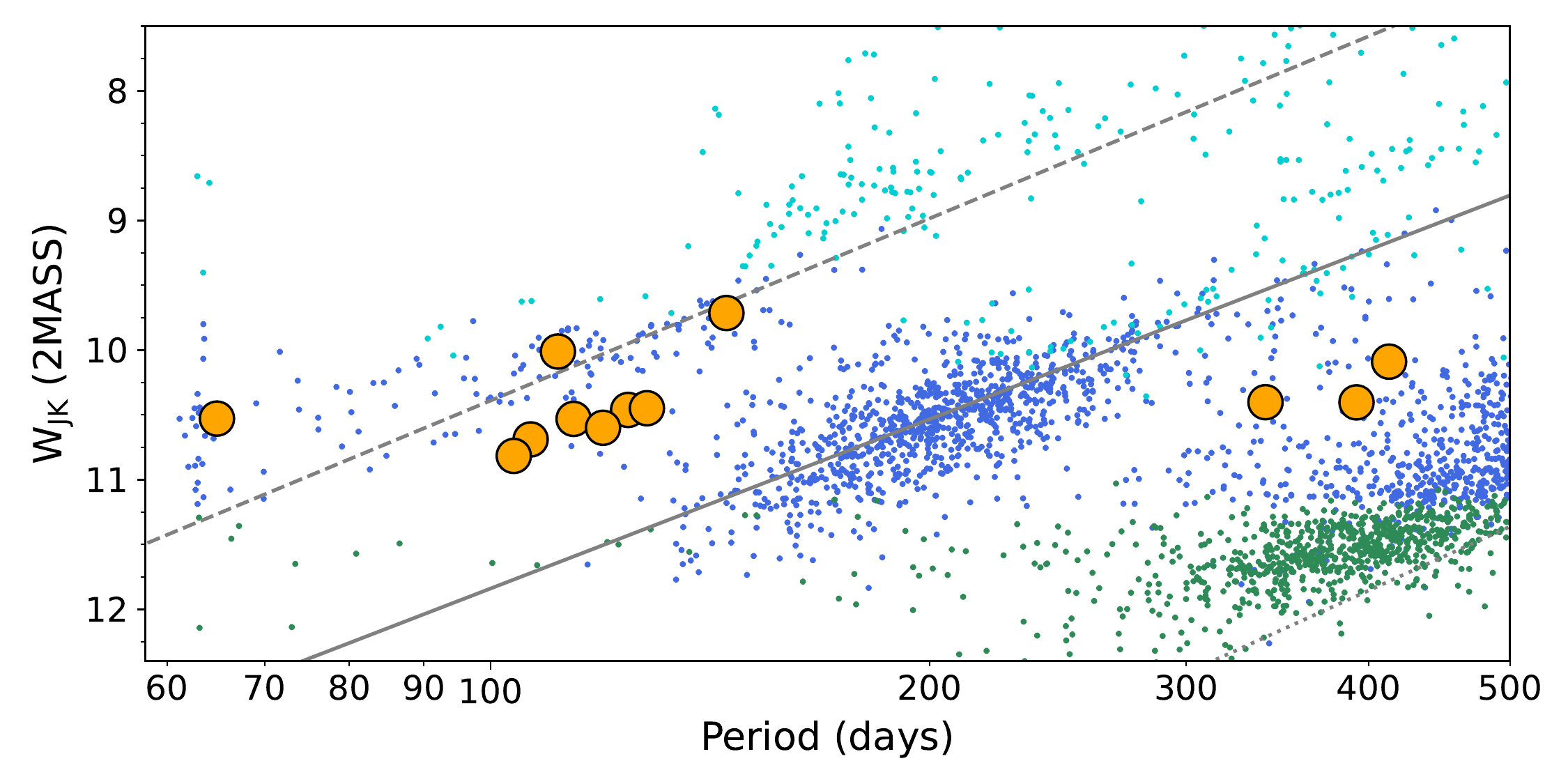}
 \caption{Period - magnitude diagram for the stars located above the Galactic plane. The PL sequences for first-overtone (dashed line), fundamental mode (solid line) and LSP (dotted line) LPV stars in the LMC are included. The distribution of LPV stars in the LMC are shown as blue, cyan and green symbols as in Figure~\ref{fig:PL_relations}.}\label{fig:PL_BPstars}
\end{figure}

Most of these stars do, however, have periods available (see Table~\ref{tab:periods}) 
which can in principle be used to derive distances. Figure~\ref{fig:PL_BPstars} 
shows the position of these stars in the period - W$_{\rm J, Ks}$ diagram, as well 
as the best PL sequences for O-rich LPVs around the LMC. In contrast with the stars discussed in section \ref{sec:lmcdist}, these stars can be easily at less than 30 kpc (i.e., be bright stars pulsating 
in the fundamental mode) and no assumptions can be made regarding their pulsation 
modes. We therefore report the two most likely distances (based on how close each star 
is to the different PL sequences) in Table~\ref{tab:BP_dist}. 

\begin{table}
	\centering
	\caption{Heliocentric distance estimates for the BP stars. The distances for two possible pulsation modes are reported. Errors on the distances are obtained by propagating the uncertainties as in Table~\ref{tab:Miras_dist}.}
	\label{tab:BP_dist}
\begin{tabular}{ccccc}
\hline
ID     & Pulsation & Distance & Pulsation & Distance\\
       & mode      & (kpc)    & mode & (kpc) \\
\hline
BP22  & FM &  74.0 $\pm$ 2.7 & LSP & 24.2 $\pm$ 0.9\\
BP24  & FO &  60.4 $\pm$ 1.5 & FM  & 30.8 $\pm$ 0.9\\
BP25  & FO &  62.8 $\pm$ 1.6 & FM  & 31.7 $\pm$ 0.9\\
BP26  & FM &  83.8 $\pm$ 3.2 & LSP & 27.6 $\pm$ 1.1\\
BP27  & FO &  59.8 $\pm$ 1.4 & FM  & 30.4 $\pm$ 0.8\\
BP32  & FO &  62.5 $\pm$ 1.5 & FM  & 32.0 $\pm$ 0.9\\
BP33  & FM &  75.9 $\pm$ 2.9 & LSP & 25.1 $\pm$ 1.0\\
BP35  & FO &  46.0 $\pm$ 1.1 & FM  & 23.4 $\pm$ 0.7\\
BP36  & FO &  51.4 $\pm$ 1.3 & FM  & 25.7 $\pm$ 0.8\\
BP37  & FO &  64.2 $\pm$ 1.5 & FM  & 32.4 $\pm$ 0.9\\
BP38  & FO &  35.3 $\pm$ 0.9 & FM  & 18.6 $\pm$ 0.6\\
BP39  & FO &  64.5 $\pm$ 1.6 & FM  & 32.7 $\pm$ 1.0\\
\hline
\end{tabular}
\end{table}

Given the uncertain distances for these stars, and their proper motions 
centered on $(\mu_{\alpha}, \mu_{\delta}$) = (0, 0) mas yr$^{-1}$, we cannot 
rule out the possibility of them being distant ($\sim$ 30 kpc) Galactic Mira-like stars, rather 
than being associated with the LMC or SMC. This group of stars \textit{is} located 
close to the predicted area of the sky where tidally-stripped LMC RRL  
stars have been reported by \cite{Petersen22}. However, based on the mock observations 
presented in \citeauthor{Petersen22}, LMC debris is expected to have much larger 
line-of-sight velocities than that measured for the BP stars (V$_{\rm los} >$ 300 km 
s$^{-1}$). Further investigation to understand the origin of the BP stars (as likely distant Galactic 
Mira-like stars) is needed, however it is beyond the scope of this work. Accordingly, for the remainder of this paper, we will focus on the Mira-candidates in the vicinity 
of the Clouds. 

\section{Outer periphery of the LMC}\label{sec:discussion}

\begin{figure*}
\centering
 \includegraphics[width=0.8\textwidth]{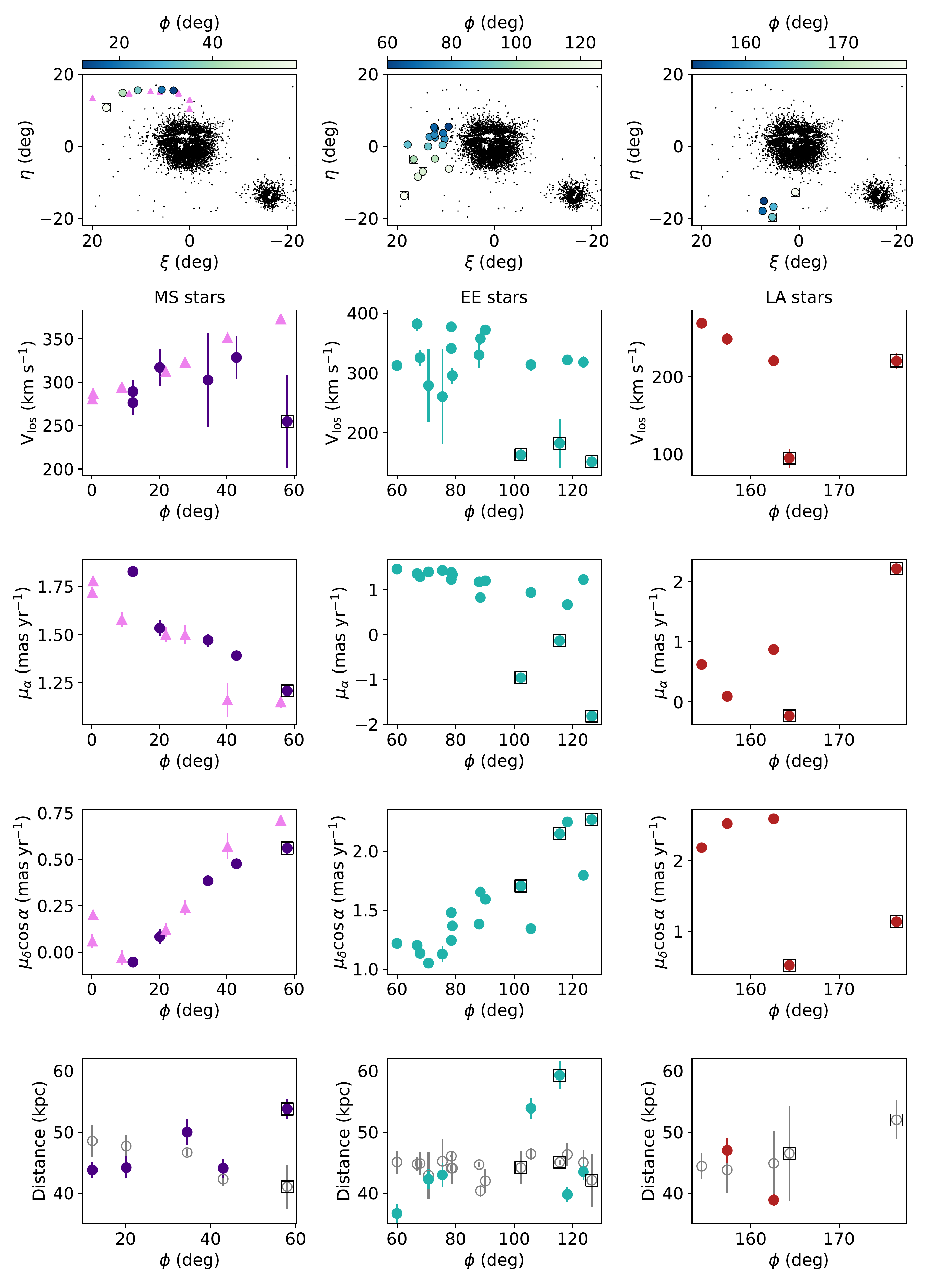}
 \caption{Phase-space properties for our targets. The left, middle and right panels correspond to stars in the MS, EE and LA groups. The top panels show the on-sky distribution of the Mira candidates selected by \protect\cite{deason17}. In the left panels, the violet triangles are measurements from \protect\cite{Cullinane22} for different fields along the LMC northern arm. Distances derived based on PL relations are shown as coloured circles, while those corresponding to the mean distance of the closest RRL stars are shown as grey open circles. Possible kinematic outliers for each group are denoted with open square outlines in each panel.} \label{fig:observables}
\end{figure*}

Figure~\ref{fig:observables} shows the phase-space information available 
for the targets in the three different groups of stars in the Magellanic 
periphery. The on-sky position as well as radial velocities, proper 
motions and distances, are shown. The angle $\phi$ is used as a position 
angle, going from 0 to 180{\degree} anticlockwise from North toward South. 
In the bottom row, the mean distance of the closest RRL is presented as a grey symbol. 
In this figure, it is possible to recognise trends and isolate potential 
outliers for each group. 

In the case of the MS stars (left panels), 
the target at the largest on-sky distance with respect to the LMC center 
(MS02, $\phi$ = 60{\degree}) has a rather smaller radial velocity and 
potentially deviating distance compared to the rest of the stars in 
the group. As a comparison, the left panels include the measurements 
presented by \cite{Cullinane22} for different fields along the 
northern arm (violet crosses). It is clear that our targets trace 
the same trends in radial velocity and proper motions as in 
\cite{Cullinane22}, with MS02 most likely being an outlier. The trend in distances, being those 
stars at large $\phi$ slightly more distant than those close to the 
LMC center, can be explained considering the inclination and orientation 
of the line of nodes of the LMC plane, as already discussed in \cite{Cullinane22}.

For the LA stars (right panels), there seems to be one outlier based on its 
rather small radial velocity and proper motions (LA43, $\phi$ = 165{\degree}). 
LA 44, at $\phi$ = 176{\degree}, also has a deviating proper motion compared to 
the rest of the stars in the group. We therefore consider both as 
potential contaminants.

In the case of EE stars, most of the stars have large ($>$ 300 km s$^{-1}$) 
velocities, except for three stars: EE17, EE18 and EE22. These stars are located at 
the largest on-sky distances from the LMC center. Their proper motions, 
particularly $\mu_{\alpha^{\ast}}$, also deviate from the other 
stars in the group. We decided to flag them as potential outliers, 
although it could be that these stars are instead tracing a different structure, 
as they all have very similar, relatively large velocities ($>$ 150 km s$^{-1}$).

\subsection{3D motions}

With 6D phase-space information in hand, the 3D motion for each star in the LMC reference frame can be calculated. In particular, 
the cylindrical velocities (V$_R$, V$_{\phi}$ and V$_Z$) can be used to assess the origin (e.g., disturbed LMC disk stars) 
of our targets. Based on the \cite{vanderMarel01, vanderMarel02} formalism, the radial motion V$_R$, azimuthal motion V$_{\phi}$ and the 
velocity perpendicular to the disk plane V$_Z$, can be estimated as 

\begin{eqnarray}
    V_R = \left(x' v_x' + y' v_y'\right)/R \text{,}\\ \nonumber
    V_{\phi} = \left(y' v_x' - x' v_y'\right)/R \text{,}\\\nonumber
    V_Z = v_z'\textbf{.}
\end{eqnarray}

$R$ is the in-plane radial distance of a tracer to the LMC center, while (v$_x'$, v$_y'$, v$_z'$) are the Cartesian velocities in the plane of the LMC disk, after subtracting the LMC systemic motion. To derive these Cartesian velocities, the formalism derived by \cite{vanderMarel01, vanderMarel02} was followed. The LMC center-of-motion (COM) was fixed at ($\alpha_{0}$, $\delta_0$) = (79\fdg88, --69\fdg59) \citep{vanderMarel14}, while the line-of-sight velocity and the proper motions of the COM used were V$_{\rm los}$ = 261.1 km s$^{-1}$, and ($\mu_{\alpha} \cos{\delta}$, $\mu_{\delta}$) = (1.895, 0.287) mas yr$^{-1}$, from \cite{vanderMarel14}. The adopted LMC heliocentric distance was $D_{\rm H}$ = 49.5 kpc \citep{pietrzynski19}. For the LMC disk geometry, the values reported by \cite{Choi18} were used, following the discussion in \cite{Cullinane22}: i.e., $i$ = 25\fdg86 and $\theta$ = 239\fdg23 respectively.

\begin{figure}
\centering
 \includegraphics[width=0.47\textwidth]{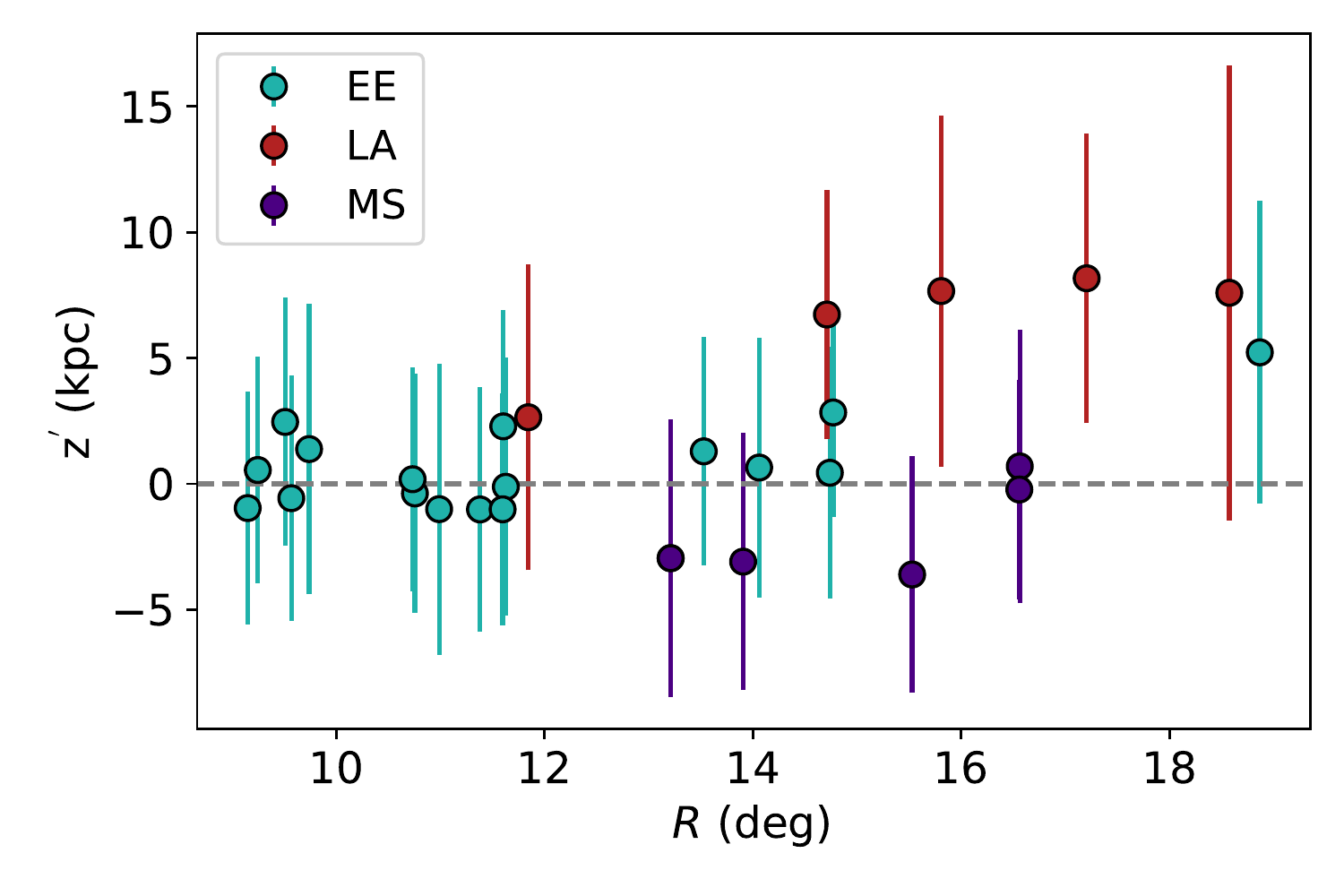}
 \caption{In-plane radial distance $R$ versus vertical distance z$^{'}$ for the stars in the vicinity of the LMC. Beyond $\sim$15{\degree}, the stars are far from the LMC disk plane.} \label{fig:Zplane}
\end{figure}

We adopted the distances of the closest RRL stars in the sky as a first rough estimate of the distance of each Mira candidate, even for those without periods available in the literature. As shown in Fig.~\ref{fig:RRL_distances}, for those stars with available periods, the distance derived using PL-relations for long-period variables in the LMC and those obtained based on the closest RRL stars are compatible. To test the influence of the distance determination used, we also computed the cylindrical velocities assuming that all our targets are on the LMC disk plane (z' = 0 kpc), and therefore their on-plane distance is derived as

\begin{equation}
    D_{\rm disk} = \frac{D_0 \cos{i}}{\cos{i} \cos{\rho} - \sin{i}\sin{\rho}\sin{(\phi-\theta)}},
\end{equation}

where $(\rho, \phi)$ are the on-sky distance and position angle of our targets respectively. The cylindrical velocities obtained this way are, for most of the stars, compatible within the errors with those values derived adopting the RRL-based distance. However, for the stars at a large projected radius $R$, the vertical position $z^{'}$ could significantly deviate from the disk plane. Figure~\ref{fig:Zplane} shows the in-plane projected radius $R$ of each star and the corresponding vertical position z$^{'}$, estimated based on the RRL-based distance, which is positive for stars above the LMC disk (in the direction towards the observer). Stars located at $R >$ 14{\degree} are located more than 5 kpc away from the LMC disk, in agreement with the results for red-clump stars presented in \cite{Cullinane22b}. Therefore, we adopted the individual distances for each star instead of assuming that all of them are located in the LMC disk plane. The gnomonic-projected coordinates ($\eta$, $\xi$) and 3D velocities are presented in Table~\ref{tab:VrVphiVz}, including the corresponding errors derived from the propagation of the uncertainties in the measured proper motions, radial velocities and distances.

\begin{figure*}
\centering
 \includegraphics[width=\textwidth]{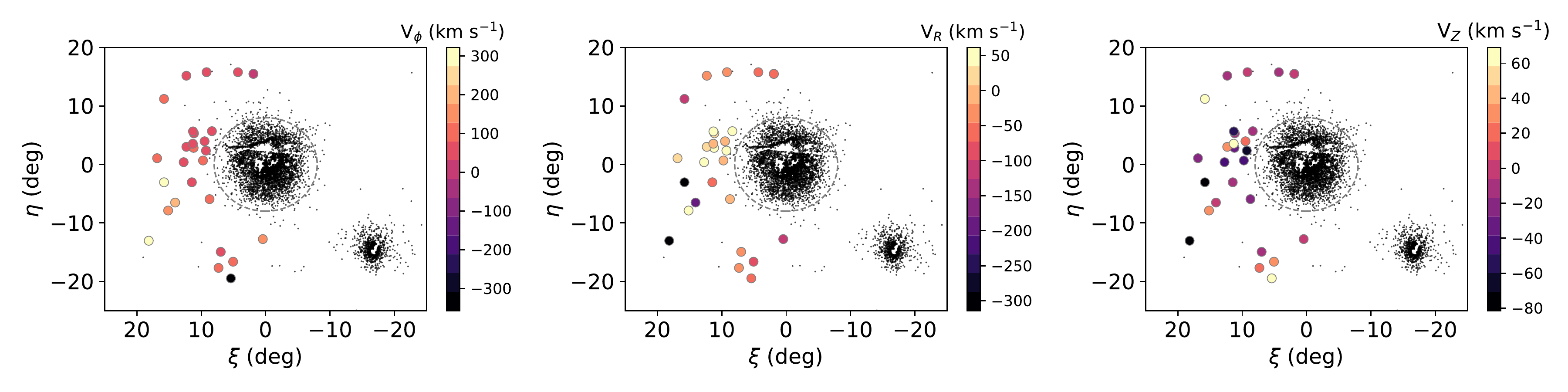}
 \caption{The azimuthal velocity (V$_{\phi}$), radial velocity (V$_{R}$), and vertical velocity (V$_{\rm Z}$) for our targets. The dashed circle marks 9{\degree} from the LMC center. The black points are the Mira candidates from \protect\cite{deason17}} \label{fig:cylindric}
\end{figure*}

\begin{table}
	\centering
	\caption{3D cylindrical velocities in the LMC coordinate system. }
	\label{tab:VrVphiVz}
\begin{tabular}{rrrccc}
\hline
ID     & $\xi$ & $\eta$ &  V$_{\phi}$   & V$_{R}$       & V$_{Z}$       \\
       & (deg) & (deg)    & (km s$^{-1}$) & (km s$^{-1}$) & (km s$^{-1}$) \\
\hline
EE06 & 15.2 &  --7.9 &   135.9 $\pm$  5.5 &    43.9 $\pm$  6.0 &   33.8 $\pm$  6.6 \\
EE07 & 11.5 &  --3.0 &    33.3 $\pm$  7.8 &  --48.5 $\pm$  7.9 & --13.1 $\pm$  9.4 \\
EE08 & 12.7 &    0.4 &    48.1 $\pm$  8.2 &    37.1 $\pm$  8.2 & --39.9 $\pm$  7.9 \\
EE09 &  9.3 &    2.4 &    38.4 $\pm$  8.2 &    39.6 $\pm$  8.7 & --60.8 $\pm$  7.9 \\
EE10 &  8.7 &  --5.9 &    85.2 $\pm$  6.8 &   --5.3 $\pm$  7.8 & --26.8 $\pm$  9.0 \\
EE11 &  9.7 &    0.7 &    81.5 $\pm$ 11.9 &     0.4 $\pm$  8.7 & --38.6 $\pm$ 18.9 \\
EE12 & 11.2 &    2.8 &    86.5 $\pm$  8.2 &    44.1 $\pm$  8.1 & --29.2 $\pm$  7.9 \\
EE13 & 12.3 &    3.0 &    68.2 $\pm$  8.8 &    23.6 $\pm$  7.5 &   36.5 $\pm$ 12.1 \\
EE14 & 11.3 &    3.5 &    55.0 $\pm$ 39.3 &   --3.8 $\pm$ 21.7 &   69.0 $\pm$ 70.2 \\
EE15 & 11.2 &    5.3 &    66.9 $\pm$  8.1 &    11.5 $\pm$  6.5 & --7.9  $\pm$ 12.2 \\
EE16 &  9.5 &    4.0 &    72.7 $\pm$ 27.6 &  --12.4 $\pm$ 12.2 &   22.6 $\pm$ 54.1 \\
EE17 & 14.1 &  --6.5 &   222.5 $\pm$ 14.7 & --189.2 $\pm$ 21.5 &    4.2 $\pm$ 32.7 \\
EE18 & 15.8 &  --3.1 &   321.5 $\pm$  6.8 & --305.5 $\pm$  6.4 & --78.6 $\pm$  8.9 \\
EE19 &  8.4 &    5.7 &    78.6 $\pm$  7.8 &    61.6 $\pm$  7.5 & --12.9 $\pm$  7.9 \\
EE20 & 16.9 &    1.1 &    82.6 $\pm$  7.9 &    10.2 $\pm$  7.0 & --23.4 $\pm$  9.2 \\
EE21 & 11.3 &    5.6 &    39.4 $\pm$  7.7 &    43.5 $\pm$  6.8 & --49.6 $\pm$ 10.1 \\
EE22 & 18.2 & --13.0 &   283.1 $\pm$  5.4 & --314.1 $\pm$  6.8 & --81.4 $\pm$  7.1 \\
LA40 &  7.3 & --17.7 &    81.9 $\pm$  5.7 &  --41.2 $\pm$  6.4 &   20.5 $\pm$  6.9 \\
LA41 &  5.1 & --16.6 &   122.1 $\pm$  6.5 &  --73.0 $\pm$  6.6 &   34.7 $\pm$  6.8 \\
LA42 &  7.0 & --14.9 &    43.8 $\pm$  6.7 &  --40.2 $\pm$  6.7 &  --9.1 $\pm$  6.9 \\
LA43 &  5.4 & --19.5 & --357.4 $\pm$  6.5 &  --52.0 $\pm$  8.0 &   64.1 $\pm$  9.9 \\
LA44 &  0.4 & --12.8 &   133.4 $\pm$  7.8 & --102.6 $\pm$  8.0 &  --1.5 $\pm$  9.9 \\
MS01 &  4.3 &   15.8 &    51.9 $\pm$ 15.8 &  --48.5 $\pm$ 15.6 & --16.1 $\pm$ 51.1 \\
MS02 & 15.8 &   11.2 &    96.5 $\pm$ 22.1 &  --99.8 $\pm$ 18.1 &   64.5 $\pm$ 45.8 \\
MS03 & 12.3 &   15.2 &    67.1 $\pm$  9.7 &  --40.8 $\pm$  9.6 &  --6.6 $\pm$ 22.1 \\
MS04 &  1.9 &   15.5 &   --2.9 $\pm$  6.0 &  --50.1 $\pm$  6.0 &    2.3 $\pm$ 11.1 \\
MS05 &  9.2 &   15.8 &    47.7 $\pm$  9.2 &  --25.4 $\pm$  9.0 &  --0.4 $\pm$ 19.5 \\
\hline
\end{tabular}
\end{table}

Fig.~\ref{fig:cylindric} shows the $V_{\phi}$ (left panel), V$_{R}$ (middle panel) and V$_{Z}$ (right panel) for all our targets. It is evident that the largest variations in velocity are found at the largest angular distances from the LMC (i.e., outer EE and LA stars). The azimuthal velocity is consistent with V$_{\phi}$ $\sim$ 70 km s$^{-1}$, which is the constant value reached by the LMC rotation curve \citep[see e.g.,][]{Wan2020, Gaia2021, Cullinane22b} for most of the stars, particularly those in the northern arm. A few stars have excessively large azimuthal velocities, of the order of 300 km s$^{-1}$, corresponding to those with line-of-sight velocities $<$ 200 km s$^{-1}$, while one of the LA stars has a counter-rotating motion. These deviating azimuthal velocities could be due to perturbations in the outer LMC.

The in-plane radial velocity, V$_{R}$, mildly deviates from a disk in equilibrium for several stars. In the case of those stars in the northern arm (MS), an inward motion of $\sim -$46 km s$^{-1}$ is found, consistent with the measurements of \cite{Cheng22} and \cite{Cullinane22}. For the outer EE stars and LA stars, however, V$_{R}$ reaches much larger values, up to $-$250 km s$^{-1}$, reinforcing the idea of these being perturbed LMC disk material. Large inward velocities were also reported by \cite{Cheng22} towards the south of the LMC.

The possible outlier in the MS group, MS02 (see Fig.~\ref{fig:observables}), is located slightly off from the track of the northern arc, inside the so-called "North-East Structure" \citep[NES,][]{Gatto}. Based on proper motions from {\it Gaia}, the NES was found to have in-plane velocities V$_{R}$, V$_{\phi}$ similar to the stars in the northern arm, suggesting a possible common origin. We found, however, that the out-of-plane velocity V$_{Z}$ strongly deviates from the values for the rest of MS stars, which are consistent with a disk in equilibrium. 

The out-of-plane velocities, V$_{Z}$, of other stars largely deviate from a disk in equilibrium, reaching up to $-$80 km s$^{-1}$ (EE22). Similar to the results in \cite{Cullinane22}, the out-of-plane velocities for the stars in the northern arm appear to be out of equilibrium. However, owing to the large errors in radial velocities for the MS stars, the error on V$_{Z}$ does not allow us to confirm this behaviour. As those stars in the N1 and N2 fields of \cite{Cheng22}, the radial and rotational motion of these stars is compatible with a disk origin. \cite{Cullinane22} found consistent velocities in the seven fields tracing the northern arm (see dashed blue circles in Fig.~\ref{fig:Mira_targets}), finding that the stars in them have a mean while V$_{\phi}$ of the order of $\sim$70 km s$^{-1}$, consistent with the rotation curve of the LMC \citep[see e.g.,][]{Wan2020, Gaia2021}. On contrast, V$_{R}$ and V$_{Z}$ are deviated from zero, expected for a disk in equilibrium. The five MS stars have an inward radial motion of $\sim$40 km s$^{-1}$, similar to the value derived in \cite{Cullinane22}. The in-plane vertical motion shows larger deviations, reaching up to 135 km s$^{-1}$. Discarding this measurement, the other four stars have a mean V$_Z$ of 32 km s$^{-1}$. Similar values are found in the APOGEE fields analyzed by \cite{Cheng22}. We can therefore conclude that all MS Mira candidates are consistent with LMC perturbed disk material.

The 3D velocities of the EE stars in the northern-east area ($\eta$, $\xi >$ 0{\degree}) are consistent with a perturbed disk, having a median azimuthal velocity of $V_{\phi} \sim$ 65 km s$^{-1}$, in-plane radial velocities V$_{R}$ going from almost zero up to $\sim$60 km s$^{-1}$, and negative out-of-plane velocities. These relatively large in-plane radial and out-of-plane velocities are contrary to the findings of \cite{Cullinane22b}, based on the aggregate motion of red clump stars in three fields at the north-east of the LMC, which were found to be consistent with a disk in equilibrium. This apparent disagreement can potentially be due to the larger angular separations from the LMC center traced by the EE stars (from to 16{\degree} up to $\sim$21{\degree}) compared to the fields in \cite{Cullinane22} (see Fig.~\ref{fig:Mira_targets}). 

For the EE stars in the southern-east region, the azimuthal and in-plane radial velocities are exceedingly large, reaching out to $V_{\phi} \gtrsim$ 300 km s$^{-1}$ and V$_{R} \sim$ --300 km s$^{-1}$. These large 3D velocities can be explained based on the different line-of-sight velocities and proper motions of these stars, see the middle panels in Figure~\ref{fig:observables}. The three potential outliers have relatively large line-of-sight velocities ($>$ 150 km s$^{-1}$) but are considerably lower than the rest of the stars in the EE group. The proper motions in the R.A. direction also strongly deviate from the values for stars in the area of sky which have smaller angular distances. We therefore cannot rule out the possibility of these stars being extremely disturbed LMC disk stars. The other EE stars in the south-east have very different 3D velocities, reflecting the fact that towards the south, the LMC disk is kinematically disturbed \citep{Cheng22, Cullinane22b}.

In the south, stars from the LA group also show deviations from a disk in equilibrium. LA43 and LA44, previously identified as potential outliers based on their line-of-sight velocities and/or proper motions, have extreme $V_{\phi}$ velocities. LA43 is found to be counter-rotating with respect to the LMC disk, while moving inwards the LMC, with V$_{R} \sim$ -- 50 km s$^{-1}$. Stars from the LA group are located towards the same direction in which \cite{Olsen11} found potential SMC debris, with apparent counter-rotating line-of-sight velocities. The CaT metallicity derived for LA43 is [Fe/H] $=-$1.42, in fair agreement with CaT metallicities for SMC stars. There are no spectroscopic studies in the literature with observations at angular distances $>$ 18\degree in the southern area around the LMC, and therefore we cannot compare this result with previous observations. 

The rest of the stars in the LA group are placed close to the southern arm-like feature reported in \cite{BelokurovErkal19}. LA42 has 3D kinematics consistent with a disk in equilibrium, although the vertical distance is $\sim$8 kpc above the LMC disk plane. These results are in very good agreement with the measurements for field 26 from \cite{Cullinane22b}, which is very close to LA42 on the sky. LA40 and LA41 have negative inward velocities and positive out-of-plane velocities, consistent with perturbed disk material, in contrast to the relatively unperturbed inner southern LMC disk reported in \cite{Cullinane22b}, and in better agreement with the results from \cite{Cheng22} for the southern periphery of the LMC.

\section{Comparison with simulations} \label{sec:sims}

\subsection{Simulation setup}

In order to explore how this dataset constrains the LMC-SMC interaction history, we produced a suite of numerical simulations. This suite is based on the simulations in \cite{BelokurovErkal19,Cullinane22,Cullinane22b}. In particular, we model the interaction of the LMC-SMC in the presence of the Milky Way. We model the LMC with a Hernquist \citep{Hernquist1990} dark matter halo and an exponential stellar disk. For the Hernquist profile, we use a mass of $1.5\times10^{11} M_\odot$, motivated by the results of \cite{Erkal2019}, and a scale radius of 20 kpc. This scale radius is chosen to match the circular velocity of the LMC measured at 8.7 kpc \citep{vanderMarel14}. For the exponential disk, we use a mass of $2\times10{9} M_\odot$, a scale radius of 1.5 kpc, and a scale height of 0.4 kpc. As in \cite{Cullinane22,Cullinane22b}, we simulate the exponential disk with tracer particles and model the gravitational potential of the LMC with a particle sourcing the combined Hernquist and exponential disk potentials. We initialize the disk with $10^7$ particles using \textsc{agama} \citep{vasiliev2019} but only simulate the $\sim$2500000
particles which have apocenters larger than 7 kpc since our dataset is focused on the outer LMC. 

The potential of the SMC is modelled as a logarithmic potential (i.e. with a flat rotation curve) within 2.9 kpc with a circular velocity of 60 km/s, based on the observations in \citep{Stanimirovic2004}. Beyond 2.9 kpc, the SMC is modelled as a Hernquist profile with a mass of $2.5\times10^9 M_\odot$ and a scale radius of 0.043 kpc. This scale radius is chosen to match the observed circular velocity at 2.9 kpc. We note that we model the inner regions of the SMC as a logarithmic potential to avoid any unphysically large perturbations during its close encounter with the LMC disk. During the simulation, the SMC is modelled as a particle sourcing this potential.

The Milky Way is modelled as a 3-component system with an NFW dark matter halo, a Hernquist bulge, and a Miyammoto-Nagai disk based on the \textsc{mwpotential2014} model in \cite{galpy}. The NFW halo has a mass of $8\times10^{11} M_\odot$, a scale radius of 16 kpc, and a concentration of 15.3. The Miyamoto-Nagai disk has a mass of $6.8\times10^{10} M_\odot$, a scale radius of 3 kpc, and a scale height of 0.28 kpc. The Hernquist bulge has a mass of $5\times10^9 M_\odot$ and a scale radius of 0.5 kpc. During the simulation, the Milky Way is allowed to move in response to the LMC and SMC. This is done by treating the Milky Way as a single particle which sources its three-component potential. 

During each simulation, the LMC and SMC are initialized given their present-day position and velocity with the Milky Way placed at the origin. The Milky Way, LMC, and SMC are rewound for 2 Gyr. At this time, the tracer particles representing the LMC disk are injected and the simulation is evolved to the present day. Since the tracer particles have a range of orbital timescales with respect to the LMC, the tracer particles are individually evolved to the present day for computational efficiency. 

Our suite consists of four sets of 100 simulations of the LMC-SMC-Milky Way encounter which differ on the present-day LMC and SMC positions and velocities. For the first set, we sampled the LMC and SMC position and velocity based on observations of their proper motions, radial velocities, and distances from \cite{Kallivayalil+13,vanderMarel02,Harris+06,pietrzynski19,Graczyk+14} respectively. We then simulated 100 realizations of the LMC and SMC's present-day positions and velocities and simulated each of these. Interestingly, in 51 of these realizations, the SMC has a negligible effect on the LMC disk since it has no close crossings with the LMC disk in the past, and thus the majority of the simulations in this suite were essentially unperturbed. For the second suite of 100 simulations, we again sampled from the LMC and SMC's present-day positions and velocities. However, for each sample, we then integrated the LMC-SMC-Milky Way orbit for 2 Gyr and required that there was at least one LMC disk crossing more ancient than 250 Myr ago. 

For the third set of simulations in the suite, we explored a slightly larger range of LMC-SMC interactions by considering larger errors on the present-day position and velocity of the LMC and SMC. This was done by accounting for the systematic uncertainty on the LMC and SMC proper motions and by accounting for the uncertainty in the on-sky location of the SMC and LMC. For the LMC, we used proper motions of $(\mu_\alpha^*,\mu_\delta)=(-1.865\pm0.015,0.331\pm0.049)$ mas yr$^{-1}$ and $(\alpha,\delta) = (80\fdg440\pm0\fdg725,-69\fdg238\pm0\fdg243$). The proper motions come from Gaussian fits to the proper motion measurements in \cite{Kallivayalil+13,Wan2020,Gaia2021,choi22,Niederhofer+2022}. The on-sky location comes from a Gaussian fit to the centers measured in \cite{vanderMarel02,vanderMarel14,Wan2020,Gaia2021,choi22,Niederhofer+2022}. For the SMC we used a proper motion of $(\mu_\alpha^*,\mu_\delta)=(-0.734\pm0.017,-1.227\pm0.010)$ mas yr$^{-1}$ and $(\alpha,\delta) = (14\fdg255\pm1\fdg713,-71\fdg710\pm0\fdg338$). For the proper motions we use Gaussian fits to the proper motion measurements in \cite{Kallivayalil+13,deleo+2020,Niederhofer+2021}. For the on-sky location we use Gaussian fits to the measurements in \cite{deVaucoleurs+72,Ripepi+17,Kallivayalil+13,diteodoro+2019}. As with the second set in the suite, we required that each of the 100 realizations had at least one SMC-LMC disk crossing more ancient than 250 Myr ago.

For the final suite of 100 simulations, we found a particular realization which was a good match to the data (see Sec.~\ref{sec:sim_vs_data}) and simulated 100 realizations with similar phase-space coordinates for the LMC and SMC. This was done by using k-Nearest Neighbors (with $k=5$) to estimate the covariance matrix for the LMC-SMC observables. The 100 realizations were drawn from this covariance matrix. This final set of 100 simulations all had at least one LMC disk crossing more ancient than 250 Myr. 

\subsection{Comparison with data} \label{sec:sim_vs_data}

In order to compare the 6D phase-space of the model kinematics with our observations, all particles within a two-degree radius around each (R.A., Dec.) position were selected. This radius was chosen to retrieve at least ten model particles even in the case of the stars at larger distances from the LMC (i.e., those more than 15{\degree} from the LMC center). We estimated the median proper motion, line-of-sight velocity and heliocentric distance of the model particles inside this radius around each of our target stars, in each of the model realizations. Then, a log-likelihood was estimated for each of the model realizations, as 

\begin{equation}
    \log{\mathcal{L}} = \sum_{i, j} -\frac{1}{2} \log{\left( 2\pi \sigma_{\rm i,j}^2\right) } - \frac{1}{2} \frac{(m_{\rm i,j,data}-m_{\rm i,j,model})^2}{\sigma_{\rm i,j}^2} \text{,}
\end{equation}
where $i$ are the observed values for the individual stars and the corresponding values for the particles in the simulations, and $j$ the four dimensions ($\mu_{\alpha^{\ast}}$, $\mu_{\delta}$, V$_{\rm rad}$, Distance). The dispersion $\sigma_{\rm i,j}$ corresponds to the sum in quadrature of the standard deviation of each component and the errors for proper motions, radial velocities and distances.

\begin{figure}
\centering
 \includegraphics[width=0.48\textwidth]{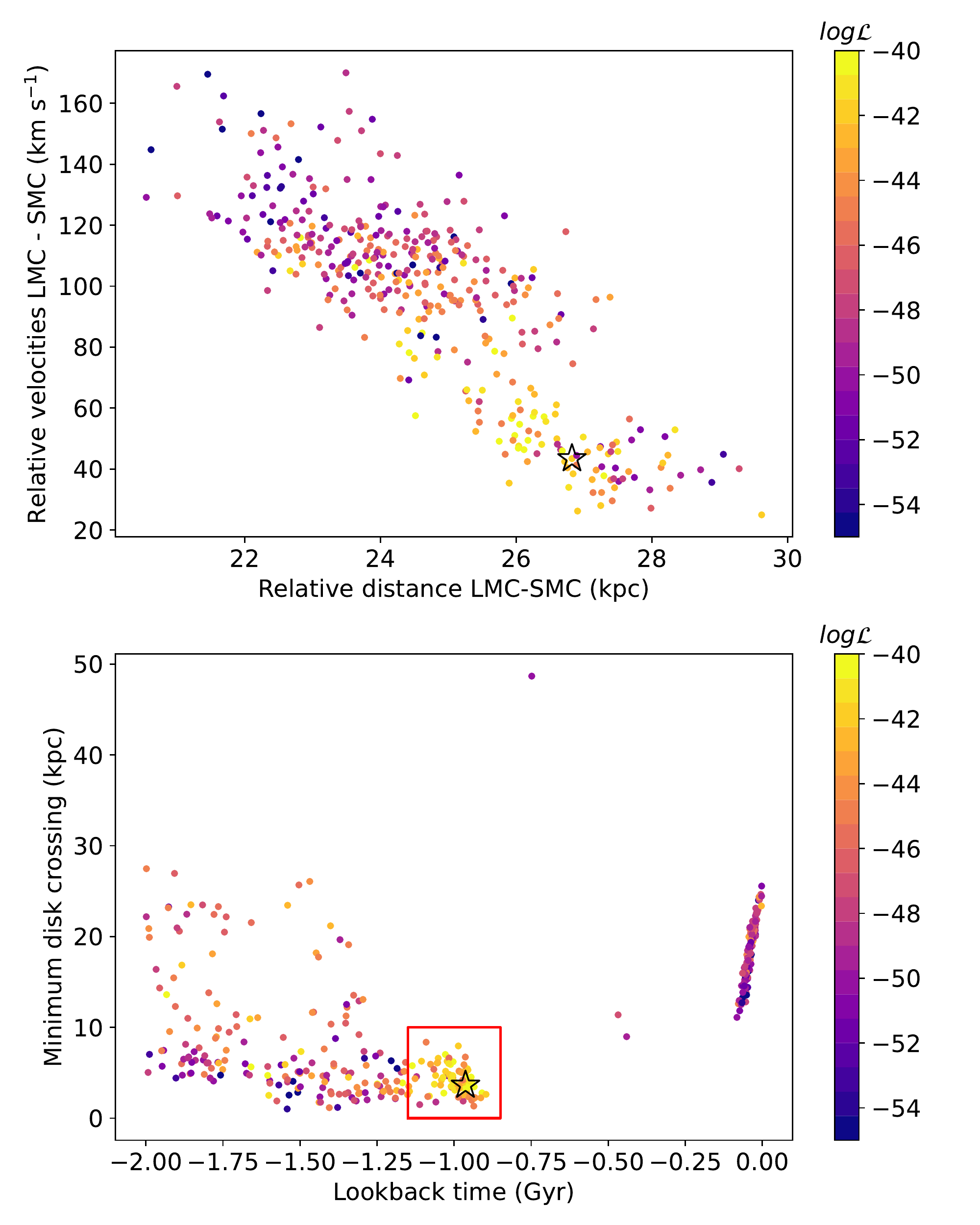}
 \caption{Top panel: present-day relative distance and velocity between the LMC and SMC in the N-body simulations. Each simulation is colour-coded according to the log-likelihood (as described in the text). Bottom panel: lookback time and the minimum radial distance $r$ of the SMC at the time it crosses the LMC's disk. The red box marks those simulations with the largest log-likelihood, i.e., those that are a better representation of our observations.}\label{fig:lbt}
\end{figure}

In the top panel of Figure~\ref{fig:lbt}, the present-day relative 3D position and velocities of the LMC and SMC, for each of the 400 model realizations, are shown, colour-coded by the log-likelihood. Those in which the present-day relative distance between the Clouds is between 24-26 kpc, with relative velocities between 80 to 40 km s$^{-1}$ have the largest $\log{\mathcal{L}}$, i.e., they best resemble the observations. The model realization that motivates the fourth suite of simulations discussed above is marked with a black star in both panels. For this comparison, the potential kinematic outliers (see Section~\ref{sec:discussion}) were not included. If they are, the results do not change significantly, and only the values for the log-likelihood decrease as the simulations are not able to reproduce the data, particularly for those stars with extreme kinematics. 

Each of the model realizations has a different interaction history between the Clouds. In particular, the SMC crosses the LMC disk between one to six times in the last 2 Gyr. The bottom panel of Figure~\ref{fig:lbt} shows the lookback time for the closest approach of the SMC to the LMC versus the radial crossing distance. Based on the log-likelihood, the simulations in which the SMC crossed the LMC's disk plane between $\sim$1.15-0.85 Gyr ago, with the SMC at a radial crossing distance $r \lesssim$ 10 kpc from the LMC have higher log-likelihood values, although there is no perfect match between any of the model realizations and our observations. All the simulations with the largest $\log{\mathcal{L}}$, inside the red box, have a very similar LMC-SMC interaction history in which the SMC crossed the LMC's disk three times in the past, at $\sim$1.78, $\sim$0.97 and $\sim$0.32 Gyr ago, at a distance of $r \sim$ 10.2, 4.5 and 5.3 kpc, respectively.

\begin{figure}
\centering
\includegraphics[width=0.48\textwidth]{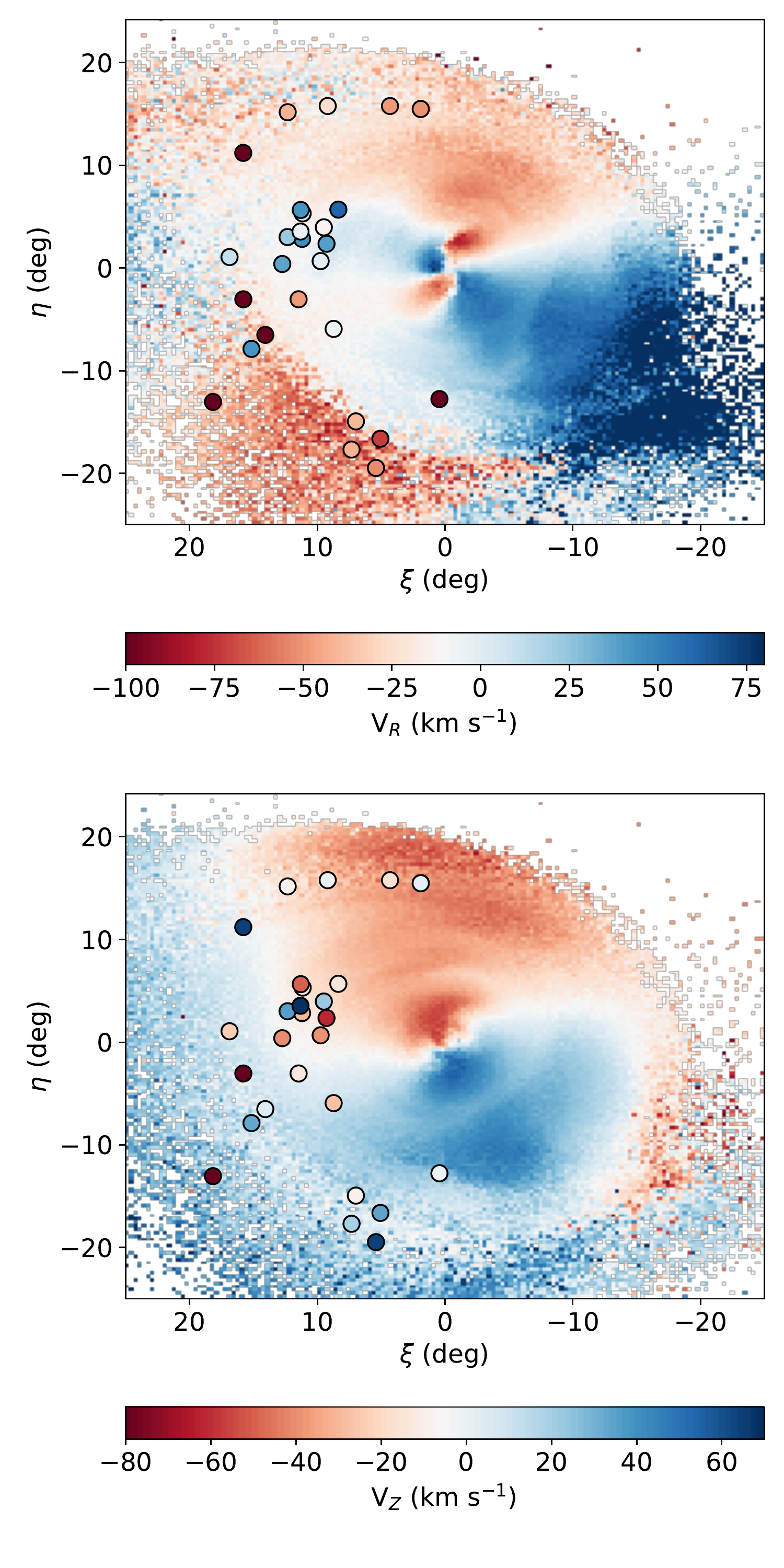}
\caption{In-plane and out-of-plane median velocities for the particles in one of the model realizations with the largest log-likelihood, i.e, closer to the 6D phase-space observations of our targets. The individual points are all the Mira-like stars in the vicinity of the Clouds, including those potential outliers (marked with open squares in Fig.~\ref{fig:observables}).}\label{fig:sim_final}
\end{figure}

Figure~\ref{fig:sim_final} shows the median in-plane radial velocity V$_{R}$ and out-of-plane vertical velocity V$_{Z}$ for one of the model realizations with the largest $\log{\mathcal{L}}$, and the corresponding observed velocities for each of our targets. A reasonable agreement, with a similar order of magnitude as those observed, is found for the in-plane radial velocities, while the vertical motions have larger discrepancies. The fact that there is no single model realization that is able to completely reproduce our 6D phase-space observations reflects the complex parameter space and interaction history of the past orbits of the Clouds. 

Based on the radial velocities, the simulation is able to reproduce the inward motion of the MS stars and those stars in the LA group (southern outer disk). In the case of the stars on the eastern side, the northern EE stars ($\eta >$ 0{\degree}) have outwards in-plane radial motions, possibly due to being pulled out from the LMC at a different time than the stars in the north or south. South-eastern stars present both inward and outward motions, which could be explained as the result of the successive SMC's disk crossings.

The vertical velocities of the MS stars are almost zero, being them on the LMC disk plane, while the potential outlier, MS02, has a larger positive vertical velocity. While in the case of EE stars, the vertical motion is less coherent, some stars having positive and others negative in plane velocities. LA stars have out-of-equilibrium V$_{z}$ velocities, similar to the median values found in this particular model realization. 

Overall, this picture in which at least two SMC's disk crossings, besides the most recent one at $\sim$150 Myr ago, is consistent with the analysis in \cite{Cullinane22b}. In that work, the Clouds orbits were traced back up to 1 Gyr ago, and older interactions were not explored. In this work, the orbits are traced up to 2 Gyr back and even though most of the model realizations have a disk crossing more than 1.3 Gyr ago, the radial crossing distance is generally much larger than $\simeq$10 kpc. There is a small number of model realizations in which the disk crossing at $\sim$ 2 Gyr has the smallest radial crossing distance (i.e., the one that most affected the LMC disk) along the past SMC orbit. Figure~\ref{fig:lbt} shows that some of them have a relatively large log-likelihood, and therefore the impact of an older disk crossing as the most important perturbation of the LMC disk cannot be fully ruled out.

Among those model realizations inside the red box in Fig.~\ref{fig:lbt}, which have high log-likelihood, we investigated the individual orbits of the SMC around the LMC. We found that all of those model realizations have the SMC crossing the LMC's disk three times, at 1.8$\pm$0.6 Gyr ago, 962$\pm$32 Myr ago, and at $\sim$320$\pm$25 Myr ago. During these three times, the SMC crossed the disk of the LMC at radial crossing distances of 10.4$\pm$2.0 kpc, 4.7$\pm$1.3 kpc and 5.4$\pm$1.5 kpc, respectively. The reported values are the mean and standard deviation values based on the individual values for each of the model realizations inside the red box in Fig.~\ref{fig:lbt}. The standard deviations of the impact parameters are relatively low, reflecting that the past interaction histories of this sub-sample of model realizations are very similar. Based on these mean values, the radial crossing distances at $\sim$950 Myr and at $\sim$300 Myr ago are very similar, and consistent with two close pericentric passages of the SMC. The mean pericentric radial and vertical distances in this sub-sample of model realizations are r$_{\rm peri}$ = 5.2$\pm$1.3 kpc, $z_{\rm peri}$ = 5.0$\pm$1.1 kpc (at $\sim$250 Myr ago), r$_{\rm peri}$ = 4.2$\pm$1.3 kpc, $z_{\rm peri}$ = --1.2$\pm$0.4 kpc (at $\sim$950 Myr ago), and r$_{\rm peri}$ = 4.6$\pm$1.0 kpc, $z_{\rm peri}$ = --2.5$\pm$0.9 kpc (at $\sim$1.81 Gyr ago). The most recent pericentric passage is consistent with previous studies \citep{choi22, Cullinane22b} in time and pericentric distance, while the vertical distance at pericenter is similar to the values reported in \cite{Cullinane22b}\footnote{The sign convention for the $z$-axis in this work is the opposite of the one adopted in \cite{Cullinane22b}.}. In both previous encounters, the pericentric distance of the SMC is on the order of $\sim$4.5 kpc, closer than the most recent interaction. 

\section{Summary and conclusions}\label{sec:conclusions}

We have carried out spectroscopic follow-up for more than 40 Mira-candidates around the LMC. Radial velocities 
obtained from low-resolution spectra, as well as stellar parameters, were derived for all the targets. Using \textit{Gaia} 
DR3 parallaxes, foreground stars among our sample were discarded.

For a subsample of the remaining stars, light curves and pulsation periods were found, confirming 
their variable nature. After classifying the stars as C-/O-rich based on the Wesenheit functions 
W$_{\rm BP, RP}$ and W$_{\rm J, Ks}$, the most likely pulsation mode was assigned based on the known 
distribution of LPVs around the LMC. Heliocentric distances based on PL relations were estimated 
for those stars with periods in the literature. We compared the distances obtained in this way with 
the median distance of RRL stars around each of the Mira-candidates, finding a reasonable agreement. We therefore adopted the RRL-based distance for the complete sample. The 6D phase-space information 
was obtained for 27 Mira candidates. For the subgroup of 14 stars located above the Galactic plane, we were 
not able to estimate a reliable heliocentric distance, although based on their proper motions these stars 
are most likely distant Galactic Miras.

A suite of N-body simulations for the LMC/SMC past interaction history (including the Milky Way) were run. 
From the phase-space properties of the simulated particles, the most likely simulations given our 
observations were identified. A scenario in which the SMC has had three close pericentric passages 
around the LMC seems to best resemble our observations. This particular interaction history is 
characterized by three LMC disk crossings of the SMC. The oldest occurred at $\sim$1.18 Gyr 
(impact parameter of 10.4 kpc), which is similar to the time at which a peak in the star 
formation history of the LMC has been found \citep{RuizLara}, and the expected time for the formation of 
the gaseous Leading Arm \citep[][]{Besla10, Diaz12}. After this, a second, much closer disk crossing took 
place at $\sim$950 Myr ago (disk crossing distance of $\sim$4.7 kpc), with this event the one that most significantly impacted the LMC periphery. A most recent disk crossing, only $\sim$350 Myr ago (disk crossing distance of $\sim$5.4 kpc) was also found to be required to recover similar kinematics to our observations. 

In summary, our observations and their comparison with the N-body simulations provide a useful 
constraint on the past interaction history of the Clouds, in which at least three close interactions 
have happened, highly disturbing the motions of the stars in the LMC periphery. Our results are similar to previous studies that claimed one or two disk crossings in the last 1 Gyr, with this work studying more distant stars (relative to the LMC center), and performing a longer integration of the Clouds's orbit (up to 2 Gyr ago). Future spectroscopic campaigns of the complete Magellanic periphery are crucial to 
obtain robust measurements of the stellar kinematics at different angular separations and position angles around the Clouds.

\section{Data Availability and online material}
All the data reduced and analyzed for the present article is fully available under reasonable request to the corresponding authors.\footnote{david.aguado@unifi.it, camila.navarrete@eso.org}

\section*{Acknowledgements}
Based on observations made at Cerro Tololo Inter-American Observatory 
at NSF's NOIRLab (NOIRLab Prop. ID: 2017B-0910; PI: M. Catelan), which is managed 
by the Association of Universities for Research in Astronomy (AURA) under 
a cooperative agreement with the National Science Foundation.
This work has made use of data from the European Space Agency (ESA) mission
{\it Gaia} (\url{https://www.cosmos.esa.int/gaia}), processed by the {\it Gaia}
Data Processing and Analysis Consortium (DPAC,
\url{https://www.cosmos.esa.int/web/gaia/dpac/consortium}). Funding for the DPAC
has been provided by national institutions, in particular the institutions
participating in the {\it Gaia} Multilateral Agreement.

JAC-B acknowledges support from FONDECYT Regular N1220083.
DA acknowledge support from the European Research Council (ERC) Starting Grant NEFERTITI H2020/808240. DA also acknowledges financial support from the Spanish Ministry of Science and Innovation (MICINN) under the 2021 Ram\'on y Cajal program MICINN RYC2021‐032609.

\bibliographystyle{mnras}

\bibliography{biblio}


\onecolumn

\begin{appendix}

\section{Gaia DR3 parameters}

Table~\ref{table_Gaia} includes the {\it Gaia} source ID, parallax, stellar parameters (T$_{\rm eff}$, log$g$, [Fe/H]) and spectral type, if available, for our targets. 

\begin{table}
	\center
	\caption{{\it Gaia} DR3 parameters for our targets. The effective temperature, surface gravity and metallicity reported are those derived from the GSP-Phot Aeneas best library using BP/RP spectra and correspond to the table columns \texttt{teff\_gspphot, logg\_gspphot, mh\_gspphot}, respectively, from the \texttt{gaiadr3.gaia\_source} table. The spectral type corresponds to the column \texttt{spectraltype\_esphs} from the \texttt{gaiadr3.astrophysical\_parameters} table.}
	\label{table_Gaia}
\begin{tabular}{rcccccr}
\hline
ID    & source ID           & parallax  & T$_{\rm eff}$ & logg          & [Fe/H]  & SpType \\ 
      &                     & (mas)     & (K)           & (cm s$^{-2}$) &         & \\ 
\hline                         
BP22  & 5460206736750537856 &    0.0323 &  4301.4  &  1.8898  &   0.1705  &  K      \\
BP23  & 3459149244108017408 &    0.0291 &          &          &           &  M      \\
BP24  & 5397474650583084160 &  --0.0207 &          &          &           &  M      \\
BP25  & 5447608326360545792 &  --0.0371 &          &          &           &  M      \\
BP26  & 5442937635326305920 &    0.0011 &  4818.8  &  2.0640  & --0.5251  &  M      \\
BP27  & 5449164032234750720 &    0.0089 &          &          &           &  M      \\
BP28  & 5672923478937609088 &    0.0155 &  33327.2 &  4.0456  &   0.0024  &  K      \\
BP29  & 5376950032669120768 &    3.2880 &          &          &           &  M      \\
BP30  & 3557444548542598912 &    3.1706 &          &          &           &  M      \\
BP31  & 3571469235968627840 &    2.0531 &  3639.7  &  4.0367  & --1.0212  &  M      \\
BP32  & 5677088493408045568 &    0.0070 &          &          &           &  K      \\
BP33  & 5670783966749167616 &    0.0361 &  4807.5  &  2.0942  & --0.5508  &  M      \\ 
BP34  & 3556135034489468672 &    2.9397 &  3724.6  &  4.5989  & --0.3160  &  M      \\
BP35  & 3483101245925708032 &    0.0134 &          &          &           &  K      \\
BP36  & 5379917099154530944 &    0.0287 &  3410.1  &  0.0619  & --0.4300  &  M      \\
BP37  & 3458916766118414592 &    0.0249 &  4406.8  &  2.0834  &   0.7920  &  K      \\
BP38  & 3477205424060762112 &    0.0086 &  4258.6  &  2.0246  &   0.2059  &  M      \\
BP39  & 3484343968942530304 &    0.0027 &          &          &           &  M      \\
\hline                                  
EE06  & 5222307398713735296 & --0.0231 &  4404.8  &  2.1391  &   0.1030   &  K     \\
EE07  & 5273716228803929088 & --0.0014 &          &          &            &  M     \\
EE08  & 5287270909369746176 & --0.0161 &  4422.5  &  2.6345  &   0.1550   &  K     \\
EE09  & 5282142924576162816 & --0.0360 &  4623.9  &  2.8845  &   0.3508   &  K     \\
EE10  & 5262926244460395520 &   0.0073 &          &          &            &  M     \\
EE11  & 5281337704108065280 &   0.0270 &  4335.3  &  1.8110  &   0.2600   &  K     \\
EE12  & 5282659076566187776 &   0.0211 &  4503.9  &  2.6225  &   0.1709   &  K     \\
EE13  & 5288705943842258304 &   0.0045 &  4347.6  &  1.8085  &   0.2000   &  K     \\
EE14  & 5285751693536837376 & --0.0171 &          &          &            &        \\
EE15  & 5286801074305659520 &   0.0301 &  4670.9  &  2.7438  &   0.1568   &  K     \\
EE16  & 5285520967893476224 & --0.0178 &          &          &            &  CSTAR \\
EE17  & 5221807017844430592 & --0.0182 &          &          &            &  M     \\
EE18  & 5272890568589181952 & --0.0102 &  4436.3  &  2.2757  &   0.1279   &  K     \\
EE19  & 5478725879815547520 & --0.0342 &  4546.5  &  2.5079  &   0.4875   &  K     \\
EE20  & 5277501920355575808 &   0.0105 &  4500.4  &  2.6607  &   0.2807   &  K     \\
EE21  & 5287009569198354816 & --0.0173 &          &          &            &  M     \\
EE22  & 5242317166011623808 &   0.0342 &          &          &            &  K     \\ 
\hline            
LA40  & 5192498745130957824 & --0.0381 &  4327.7  &  2.0829  &   0.1371   &  K     \\       
LA41  & 5194991505491879168 & --0.0063 &          &          &            &  K     \\       
LA42  & 5196276426565456128 & --0.0170 &          &          &            &  M     \\   
LA43  & 5191672668300504832 & --0.0044 &  4902.0  &  2.1046  & --1.2015   &  M     \\   
LA44  & 4621336230122857216 & --0.0028 &  5018.7  &  2.5114  & --0.7710   &  M     \\   
\hline                                                        
MS01  & 4791851791893295104 & --0.0500 &         &           &           &  M      \\   
MS02  & 5490523880100951552 & --0.0229 &         &           &           &  K      \\   
MS03  & 5498652260326236288 & --0.0514 & 4382.8  &  2.3396   &   0.7982  &  K      \\   
MS04  & 4768385293178232064 & --0.0176 & 4536.1  &  1.7935   & --0.5526  &  M      \\   
MS05  & 5499999780546395520 & --0.0473 &         &           &           &  K      \\   
\hline  
\end{tabular}
\end{table}

\end{appendix}

\label{lastpage}
\end{document}

%% file: table1_2dec.tex
\begin{table*}
	\centering
	\caption{Mira variable candidate stars observed with the COSMOS spectrograph. Columns give the equatorial coordinates of the stars, radial velocities, and stellar parameters derived from the spectra (see Section~\ref{sec:analisys}). The last column gives the quality 'Flag' for the derived stellar parameters, with {\tt flag=0} for those spectra where the spectroscopic fitting is unreliable.}
	\label{tab:stellar_params}
\begin{tabular}{rccccccccccr}
\hline
ID & R.A.           & Dec      & V$_{\rm rad}$ & $\sigma_{\rm Vrad}$ & H$_{\rm emi}$ & RV shift         & [Fe/H] & T$_{\rm eff}$ & logg          & [Fe/H]$_{\rm CaT}$ & Flag  \\
   & (hh:mm:ss.s) & (dd:mm:ss) & (km s$^{-1}$) & (km s$^{-1}$)       & ({\AA})       & (km s$^{-1}$) &        & (K)           & (cm s$^{-2}$) &                     & \\
\hline      
BP22  & 10:10:29.0 & --31:44:53 &  320.39 &  6.71 &         &         & --1.13 & 4091.71 & 0.68 & --1.05 & 1 \\
BP23  & 11:55:27.4 & --39:57:44 &  158.24 &  9.55 & 6562.25 & --25.04 & --0.91 & 3939.74 & 0.81 & --1.11 & 1 \\
BP24  & 11:17:16.7 & --36:20:11 &  324.31 &  7.60 &         &         & --1.07 & 3834.37 & 0.00 & --1.52 & 1 \\
BP25  & 10:29:52.7 & --33:29:26 &   39.59 & 10.69 & 6562.74 & --2.79  & --0.99 & 4013.93 & 0.49 & --0.98 & 1 \\
BP26  & 10:39:36.2 & --37:08:53 &  109.06 &  9.56 &         &         & --1.18 & 3747.15 & 0.00 & --0.98 & 1 \\
BP27  & 10:24:57.2 & --31:25:47 &   313.41 & 10.67 &         &         & --1.11 & 4039.59 & 0.38 & --1.26 & 1 \\
BP28  & 10:11:16.3 & --17:41:28 &  156.86 & 43.82 & 6563.25 &  20.71  & --2.93 & 3500.03 & 0.14 & --3.85 & 0 \\
BP29  & 11:21:35.3 & --43:34:16 &    0.55 &  8.65 &         &         &   0.41 & 4145.97 & 5.00 &   0.03 & 1 \\
BP30  & 10:54:28.0 & --15:59:49 & --10.04 &  9.01 &         &         & --0.22 & 4205.23 & 5.00 &   0.50 & 1 \\
      &            &            & --13.13 &  8.34 &         &         & --0.13 & 4186.30 & 5.00 &   0.50 & 1 \\
BP31  & 11:49:21.1 & --16:07:42 &  --3.22 &  9.86 &         &         & --0.96 & 3919.75 & 4.79 & --1.12 & 1 \\
BP32  & 09:30:18.0 & --21:04:54 &  234.58 & 10.70 & 6562.80 &  0.16   & --1.30 & 4037.45 & 1.10 & --1.69 & 1 \\
BP33  & 10:07:07.0 & --19:24:04 &  42.41 &  8.33 &         &         & --0.86 & 3867.57 & 0.19 & --0.98 & 1 \\
BP34  & 10:52:49.4 & --17:12:15 &   27.85 &  9.65 &         &         & --0.81 & 3967.24 & 5.00 & --0.01 & 1 \\
BP35  & 11:25:38.0 & --28:55:53 &  213.72 & 22.34 & 6562.71 & --4.06  &   0.08 & 4491.12 & 1.28 & --2.18 & 0 \\
BP36  & 11:50:27.0 & --43:22:29 &  162.73 & 86.68 &         &         & --1.63 & 3545.06 & 1.91 & --1.51 & 1 \\
BP37  & 12:03:52.8 & --40:37:26 &  235.83 & 23.03 &         &         &   0.23 & 4481.47 & 0.85 & --1.91 & 0 \\
BP38  & 11:37:54.5 & --33:58:40 &   85.07 & 12.41 &         &         & --1.01 & 3982.81 & 0.55 & --1.02 & 1 \\
BP39  & 11:45:58.0 & --27:13:38 &  160.08 &  8.87 & 6562.46 & --15.36 & --2.48 & 3500.02 & 0.00 & --2.68 & 1 \\
\hline
EE06  & 08:49:19.0 & --70:41:50 &  321.77 &  7.27 &         &         & --0.96 & 3987.77 & 0.53 & --1.00 & 1 \\
EE07  & 07:43:44.4 & --69:18:05 &  314.09 & 10.21 &         &         & --1.05 & 3833.99 & 0.24 & --1.64 & 1 \\
EE08  & 07:36:51.5 & --65:46:52 &  372.32 &  7.77 &         &         & --0.96 & 4038.42 & 0.19 & --0.96 & 1 \\
EE09  & 06:59:30.5 & --65:27:18 &  377.31 &  7.68 &         &         & --0.96 & 3958.61 & 0.03 & --0.94 & 1 \\
EE10  & 07:33:06.8 & --73:06:27 &  318.21 &  9.80 &         &         & --1.66 & 3595.92 & 0.01 & --1.07 & 1 \\
EE11  & 07:09:52.1 & --66:49:31 &  330.50 & 21.35 &         &         &   0.03 & 4496.49 & 1.28 & --0.01 & 1 \\
EE12  & 07:13:58.1 & --64:19:10 &  341.07 &  7.83 &         &         & --1.11 & 4020.17 & 0.59 & --1.10 & 1 \\
EE13  & 07:22:16.5 & --63:42:15 &  295.71 & 13.47 &         &         &   0.14 & 4499.98 & 0.80 &   0.00 & 1 \\
EE14  & 07:12:18.5 & --63:38:09 &  260.53 & 80.62 &         &         & --2.82 & 3500.01 & 2.47 & --4.78 & 0 \\
EE15  & 07:05:00.1 & --62:06:39 &  325.72 & 13.57 & 6562.29 & --23.48 & --1.62 & 4066.17 & 1.62 & --1.83 & 1 \\
EE16  & 06:56:04.7 & --63:54:41 &  279.11 & 61.28 &         &         & --2.69 & 3500.26 & 0.00 & --3.76 & 0 \\
EE17  & 08:29:56.1 & --70:29:23 &  182.25 & 41.15 &         &         & --1.82 & 3500.13 & 1.73 & --1.35 & 1 \\
EE18  & 08:19:12.4 & --66:53:22 &  162.79 & 10.26 &         &         & --0.70 & 4035.83 & 0.84 & --0.64 & 1 \\
EE19  & 06:41:52.5 & --62:38:53 &  312.61 &  7.92 &         &         & --1.16 & 3895.31 & 0.40 & --1.48 & 1 \\
EE20  & 08:03:49.1 & --63:08:30 &  357.66 & 10.26 &         &         & --0.98 & 4036.74 & 0.22 & --0.98 & 1 \\
EE21  & 07:05:05.6 & --61:45:39 &  381.97 & 10.93 &         &         & --1.34 & 3680.01 & 0.04 & --1.75 & 1 \\
EE22  & 09:55:37.2 & --71:13:39 &  150.92 &  8.40 &         &         & --0.91 & 4083.29 & 1.30 & --0.74 & 1 \\
\hline
LA40  & 09:46:56.8 & --82:16:36 &  248.70 &  7.78 &         &         & --1.04 & 3935.65 & 0.33 & --1.03 & 1 \\
LA41  & 08:42:51.7 & --83:30:38 &  220.52 &  6.90 &         &         & --1.14 & 3770.16 & 0.00 & --1.04 & 1 \\
LA42  & 08:45:03.8 & --81:04:20 &  269.10 &  7.26 & 6562.02 & --35.47 & --1.15 & 3831.63 & 1.10 & --1.87 & 1 \\
LA43  & 10:14:40.0 & --84:36:22 &   94.74 & 12.44 & 6562.79 &  --0.27 & --1.11 & 3959.30 & 0.75 & --1.42 & 1 \\
LA44  & 05:41:39.5 & --82:03:02 &  220.28 & 10.71 & 6562.30 & --22.95 & --1.19 & 3809.93 & 1.25 & --2.28 & 1 \\
\hline
MS01  & 05:57:14.3 & --53:54:02 & 317.07 & 21.15  &         &         & --0.92 & 3507.28 & 0.00 & --1.06 & 1 \\
MS02  & 07:18:21.2 & --55:19:44 & 254.90 & 53.34  &         &         &   0.46 & 4416.97 & 0.00 & --2.96 & 0 \\
MS03  & 06:48:06.4 & --53:00:33 & 328.51 & 24.66  & 6562.71 & --4.20  &   0.44 & 4441.20 & 0.00 & --2.52 & 0 \\
MS04  & 05:41:38.8 & --54:20:09 & 276.47 & 13.71  & 6562.89 &   3.95  & --1.26 & 3845.89 & 0.62 & --1.92 & 1 \\
      &            &            & 289.27 & 13.40  & 6562.74 & --2.72  & --1.25 & 3851.30 & 0.53 & --2.01 & 1 \\
MS05  & 06:28:06.1 & --53:11:05 & 302.44 & 54.11  & 6562.81 &   0.56  &   0.16 & 4468.77 & 1.26 & --3.22 & 0 \\
\hline  
\end{tabular}
\end{table*}

%% file: table2.tex
\begin{table}
	\centering
	\caption{Periods for those Mira stars with light curves available in public surveys. Gaia periods are derived from the frequency listed in the {\tt gaiadr3.vari\_long\_period\_variable} table. Periods are from either CRTS or ASAS-SN catalogues (see main text for details).}
	\label{tab:periods}
\begin{tabular}{rccc}
\hline
ID     & P$_{\rm Gaia}$ & P$_{\rm lit}$ & Source ID  \\
       & (d)            & (d)           &           \\
\hline
BP22   & 366.16     & 335.12    & SSS J101029.1-314452         \\  
BP23   & 122.99     &           &                              \\ 
BP24   & 103.14     & 106.53    & SSS J111716.7-362011         \\ 
BP25   & 121.03     & 124.26    & SSS J102952.7-332925         \\
BP26   & 407.05     & 392.42    & SSS J103936.2-370853         \\ 
BP27   & 116.71     & 114.01    & SSS J102457.2-312547         \\ 
BP32   &            & 103.73    & SSS J093018.0-210453         \\ 
BP33   & 389.43     & 413.14    & SSS J100707.0-192403         \\ 
BP35   & 110.49     & 111.22    & SSS J112538.0-285553         \\ 
BP36   & 144.37     & 145.12    & SSS J115027.0-432228         \\ 
BP37   &            & 128.00    & ASAS\_SN-V J120352.76-403726.3 \\ 
BP38   &            &  64.92    & SSS J113754.5-335839         \\
BP39   & 120.80     & 119.42    & SSS J114558.0-271337         \\ 
\hline
EE06   & 134.29     & 132.00    & ASAS\_SN-V J084919.03-704150.2 \\
EE07   & 181.09     & 180.52    & SSS J074344.4-691805         \\  
EE10   & 181.55     &           &                              \\ 
EE11   &  56.62     &           &                              \\ 
EE12   & 313.12     &           &                              \\ 
EE13   &            &   1.13    & SSS J072216.5-634214         \\ 
EE14   & 286.41     & 285.06    & SSS J071218.6-633809         \\
EE16   & 225.63     & 225.00    & ASAS\_SN-V J065604.73-635441.5 \\ 
EE17   & 473.49     &           &                              \\   
EE19   & 301.47     & 325.49    & SSS J064152.5-623852         \\   
\hline
LA40   & 423.36     &           &                              \\ 
LA41   &            &  65.88    & ASAS\_SN-V J084251.71-833038.3 \\ 
LA43   & 132.11     &           &                              \\ 
LA44   & 156.63     &           &                              \\ 
\hline
MS01   & 398.43     &           &                              \\ 
MS02   &  36.16     & 161.06    & SSS J071821.2-551943         \\  
MS03   & 158.75     & 156.39    & SSS J064806.4-530033         \\  
MS04   & 139.56     & 146.65    & ASAS\_SN-V J054138.80-542008.6 \\ 
MS05   & 215.95     & 219.39    & SSS J062806.1-531105          \\ 
\hline
\end{tabular}
\end{table}